\documentclass{vldb}
\usepackage{balance}

\newif\iftechreport
\techreporttrue

\usepackage{amsmath}
\usepackage{hyperref}
\usepackage{comment}
\usepackage{subcaption}

\usepackage{listings}
\usepackage{color}
\usepackage{comment}
\usepackage{times}

\newcommand*{\code}{\fontfamily{pcr}\selectfont}

\newtheorem{definition}{Definition}[section]

\newtheorem{lemma}[definition]{Lemma}

\newtheorem{theorem}[definition]{Theorem}

\newtheorem{example}[definition]{Example}

\definecolor{codegreen}{rgb}{0,0.6,0}
\definecolor{codegray}{rgb}{0.5,0.5,0.5}
\definecolor{codepurple}{rgb}{0.58,0,0.82}
\definecolor{backcolour}{rgb}{0.95,0.95,0.92}

\lstdefinestyle{mystyle}{
  language=java,
  morekeywords={foreach, Foreach, Node, Nbrs, OutNbrs, InNbrs, Graph, If, Sum, Graph, Node\_Prop, Nodes},
  morekeywords={Else, Count, Degree, Inf, Product, INF, Bool, Edge\_Prop, N\_P, E\_P},
  morekeywords={UpNbrs, DownNbrs, From, InBFS, InRBFS, Float, Int, Procedure},
  morekeywords={Double, OutDegree, Do, While, For, Return, InReverse, Items},
  morekeywords={InDFS, InPost, Node\_Order, Node\_Set, Node\_Seq, PushFront, 
      max=, min=, Edge, NumEdges, NumNodes, GetEdge},
  morekeywords={True,False,InDegree,NumNodes, NumEdgss, InNbrs},
  frame=lines,
  keepspaces=true,
  columns=fullflexible,
  keywordstyle=\color{blue},
  numbers=left, 
  commentstyle=\itshape,
  columns=flexible,
  xleftmargin=0.1in,
  firstnumber=auto,
  emph={ }, emphstyle=\itshape,
  escapechar=\#,
}

\usepackage{amsmath}
\usepackage{amssymb}
\usepackage{amsfonts}
\usepackage{rotating}

\usepackage{setspace}

\newcommand{\squishlist}{
  \begin{list}{$\bullet$}
   {
     \setlength{\itemsep}{0pt}
     \setlength{\parsep}{2pt}
     \setlength{\topsep}{3pt}
     \setlength{\partopsep}{0pt}
     \setlength{\leftmargin}{1em}
     \setlength{\labelwidth}{1em}
     \setlength{\labelsep}{0.5em} } }

\newcommand{\IN}{\mathrm{IN}}
\newcommand{\OUT}{\mathrm{OUT}}

\begin{document}

\title{Distributed Evaluation of Subgraph Queries Using Worst-case Optimal Low-Memory Dataflows}

\author{Khaled Ammar$^{\dag}$, Frank McSherry${\ddagger}$, Semih Salihoglu$^{\dag}$, Manas Joglekar$^{\sharp}$ \\
$^{\dag}$University of Waterloo, ${\ddagger}$ETH Z\"{u}rich,$^{\sharp}$Google, Inc\\
{khaled.ammar,semih.salihoglu}@uwaterloo.ca, frank.mcsherry@inf.ethz.ch, brahmaneya@gmail.com}

\maketitle

\begin{abstract}

We study the problem of finding and monitoring fixed-size subgraphs in a continually changing large-scale graph. We present the first approach that (i) performs worst-case optimal computation and communication, (ii) maintains a total memory footprint linear in the number of input edges, and (iii) scales down per-worker computation, communication, and memory requirements linearly as the number of workers increases, even on adversarially skewed inputs.

Our approach is based on worst-case optimal join algorithms, recast as a data-parallel dataflow computation. We describe the general algorithm and modifications that make it robust to skewed data, prove theoretical bounds on its resource requirements in  the {\em massively parallel computing} model, and implement and evaluate it on graphs containing as many as 64 billion edges.
The underlying algorithm and ideas generalize from finding and monitoring subgraphs to the more general problem of computing and maintaining relational equi-joins over dynamic relations. 

\end{abstract}

\section{Introduction}
\label{sec:intro}

{\em Subgraph queries}, i.e., finding instances of a given subgraph in a larger graph, are a fundamental computation performed by many applications and supported by many software systems that process graphs. Example applications include finding triangles and larger clique-like structures for detecting related pages in the World Wide Web~\cite{flake:web}  
and finding {\em diamonds} for recommendation algorithms in social networks~\cite{gupta:diamond}. Example systems include graph databases~\cite{neo4j, titan}, RDF engines~\cite{neumann:rdf3x, zou:gstore}, as well as many other specialized graph processing systems~\cite{aberger:emptyheaded, malewicz:pregel, teixeira:arabesque}.  As the scale of real-world graphs and the speed at which they evolve increase, applications need to evaluate subgraph queries both offline and in real-time on highly-parallel and shared-nothing distributed systems. 

This paper studies the problem of evaluating subgraph queries on large static and dynamic graphs in a distributed setting, with efficiency and scalability as primary goals. Our approach is to design distributed versions of recent worst-case join algorithms~\cite{ngo:survey, ngo:worst-case-joins, veldhuizen:leapfrog-trie-join}. We show that our algorithms require memory that is linear in the size of the input graphs and are worst-case optimal in terms of computation and communication costs (defined momentarily). We also show optimizations to balance the workload of the machines in the cluster ({\em workers} hereafter) and make our algorithms provably skew-resilient, i.e., guarantee that the costs per worker decrease linearly as we introduce additional workers.  
We prove the efficiency of our algorithms theoretically and demonstrate their practicality through extensive evaluation of their implementations in the Timely Dataflow system~\cite{murray:naiad, mcsherry:timely}. Although we focus on subgraph queries, our algorithmic and theoretical contributions apply equally to more general relational equi-joins.

\subsection{Joins and Worst-case Optimality}
Throughout the paper, we adopt the relational view of subgraph queries (as done by many previous work~\cite{aberger:emptyheaded, neumann:rdf3x, veldhuizen:leapfrog-trie-join, zeng:trinity-rdf}) in which any subgraph query can be seen as a multiway join on replicas of an {\code edge} table of the input graph. Given a directed subgraph query $Q$, we label each vertex in the query with an attribute $a_i$. 
Instances of $Q$ in an input graph $G$ is equivalent to the multiway join of tables {\code edge(a$_i$,  a$_j$)} for each edge ($a_i$, $a_j$)  in $Q$, where each {\code edge(a$_i$, a$_j$)} table contains each edge ($u$, $v$) in $G$. 
For example, the directed triangle query, in Datalog syntax, is equivalent to:
\begin{align*}
&\text{\small{{\code tri(a$_1$,a$_2$,a$_3$) := edge(a$_1$,a$_2$),edge(a$_2$,a$_3$),edge(a$_3$,a$_1$)}}}
\end{align*}

In the serial setting, a join algorithm is {\em worst-case optimal} for a query $Q$ if its computation cost is not asymptotically larger than the AGM bound of $Q$~\cite{atserias:agm}, which is the maximum possibly output size for the given size of the relations in $Q$. We refer to this quantity as $MaxOut_Q$. For example, on a graph with $\IN$ edges, $MaxOut_Q$ for the triangle query is $\IN^{3/2}$. Analogously, we say a distributed algorithm has worst-case optimal computation and communication costs, if respectively the total computation and communication done across all workers are $O(MaxOut_Q)$, for any parallelism level, i.e.,  number of workers. 

\subsection{Existing Approaches}
Existing distributed approaches that can be used to evaluate general subgraph queries can be broadly grouped into two classes: (i) edge-at-a-time approaches~\cite{choudhury:continuous-subgraph, gao:continuous-subgraph, husain:rdf, neumann:rdf3x, sun:subgraph, teixeira:arabesque,  zeng:trinity-rdf} that correspond to binary join plans in relational terms;  
and (ii) those that use variants of the Shares~\cite{afrati:mrjoins} or Hypercube~\cite{beame:communication, beame:skew, koutris:worst-case-parallel} algorithm. We also review a recent vertex-at-a-time approach that has been used in the serial setting and on which we base our distributed algorithms. 

\vspace{15pt}
\subsubsection{Edge-at-a-time Approaches}

Perhaps the most common approach to finding instances of a query subgraph is to treat it as a relational query, and to execute a sequence of binary joins to determine the result. 
For example, this approach computes: 
\begin{align*}
&\text{{\code open-tri(a$_1$,a$_2$,a$_3$):=edge(a$_1$,a$_2$),edge(a$_2$,a$_3$)}} \\
&\text{{\code tri(a$_1$,a$_2$,a$_3$):=open-tri(a$_1$,a$_2$,a$_3$),edge(a$_3$,a$_1$)}}
\end{align*}

Recent developments~\cite{ngo:survey, ngo:worst-case-joins} have shown that edge-at-a-time approaches are provably suboptimal.
For example on a graph with $\IN$ edges, any edge-at-a-time approach will do $O(\IN^2)$ computation in the worst-case and comparable communication in the distributed setting, which is worse than the AGM bound of $\IN^{3/2}$. This is because irrespective of the join order, the worst-case size of the first join is $O(\IN^2)$. 
Although couched in the language of worst-case bounds, these suboptimalities do manifest on real graph datasets, especially those demonstrating {\em skew}. For example, the largest graph we consider has a maximum degree of 45 million, and any algorithm that considers the $(4.5 \times 10^7)^2 \approx 2 \times 10^{15}$ candidate pairs of neighbors of the maximum degree vertex will simply not work.

Different systems have several optimizations on top of this basic approach including: (i) picking different join orders; (ii) decomposing the query into several subqueries; and (iii) preprocessing and indexing commonly appearing subqueries~\cite{cheng:graph-pattern-matching, he:graphql, lai:twintwigjoin, lai:seed, zou:distance-join}. None of these techniques correct the asymptotic sub-optimality. 

\subsubsection{The Shares Algorithm}
\label{subsec:shares}
The second existing technique for distributed evaluation of subgraph queries is to use the {\em Shares} of {\em Hypercube} join algorithm from references~\cite{afrati:mrjoins, beame:communication, beame:skew, koutris:worst-case-parallel}.
Consider a distributed cluster with $w$ workers and a query with $n$ relations and $m$ attributes, i.e., $n$ is the number of edges and $m$ is the number of vertices in the query subgraph.
Shares divides the $m$-dimensional output space equally over the $w$ workers and replicates each tuple $t$ of each relation to every worker that can produce an output that depends on $t$. Finally, each worker runs any local join algorithm on the inputs it receives.

There are several advantages of Shares. For most queries and parallelism levels $w$ (but not all), Shares' communication cost is less than the AGM bound (and often much less). In addition, in distributed bulk synchronous parallel systems, in which the computation is broken down into a series of {\em rounds}, Shares requires a very small number of rounds. 
However, Shares' cumulative memory requirement is $O(w^{1-\epsilon}\IN)$ and its per worker memory requirement is $O(\frac{\IN}{w^{\epsilon}})$. Here $\IN$ is the size of the input and $\epsilon \in [0, 1]$ is a query-dependent parameter. This implies a super-linear cumulative memory growth and sub-linear scaling of per-worker memory (and workload) as $w$ increases. For example, for the triangle query, $\epsilon = 1/2$. Often $\epsilon$ is much smaller, and scaling becomes an increasingly resource-inefficient way to improve performance.  
\subsubsection{Vertex-at-a-time Approaches}
\label{subsec:wcja}
In the serial setting, Ngo et. al. and soon after Veldheuizen recently developed the first worst-case optimal join algorithms called respectively the {\em NPRR}~\cite{ngo:worst-case-joins} and  {\em Leapfrog TrieJoin}~\cite{veldhuizen:leapfrog-trie-join} algorithms. These algorithms were shown to be instances of another algorithm called {\em Generic Join}~\cite{ngo:survey} (GJ), on which we base our algorithms. 
In graph terms, these algorithms adopt a {\em vertex-at-a-time} evaluation technique. Specifically, on a query that involves $a_1,..., a_m$ vertices, these algorithms first find all of the $(a_1)$ vertices that can end up in the output. Then they find all of the (a$_1$, a$_2$) vertices that can end up in the  output and so forth until the final output is constructed. When extending a partial subgraph to a new vertex a$_i$, all of the edges that are incident on a$_i$ are considered and intersected. For example, on the triangle query, these algorithms would first find all $(a_1)$ vertices and then $(a_1, a_2)$ edges that can possibly be part of a triangle. 
Then the algorithms extend these edges to ($a_1$, $a_2$, $a_3$) triangles by intersecting $a_1$'s incoming and $a_2$'s outgoing edges. Compared to edge-at-a-time approaches, these algorithms will never generate intermediate data larger $MaxOut_Q$. We note that Turbo$_{ISO}$~\cite{han:turboiso} is a serial algorithm that was developed independently in the context of subgraph matching around the same time as NPRR and Leaprfrog Triejoin. The algorithm is not worst-case-optimal but overall adopts a vertex-at-a-time approach.

\subsection{Our Approach and Contributions}
Our approach is based on recasting the basic building block of the GJ algorithm as a distributed dataflow computation primitive. 
We optimize and modify this basic primitive to obtain different algorithms tailored for different settings and achieving different theoretical guarantees. 
Our contributions are as follows:
\squishlist
\item[1.] A distributed algorithm called {\em BiGJoin} for static graphs that ach-ieves a subset of the theoretical guarantees we seek.
\item[2.] A distributed algorithm called {\em Delta-BiGJoin} for dynamic graphs that achieves the same guarantees as BiGJoin in insertion-only workloads.
\item[3.] A distributed algorithm called {\em BiGJoin-S} for static graphs that achieves all of the theoretical guarantees we seek including workload balance across distributed workers on arbitrary input instances. 
\end{list}

We implement BiGJoin and Delta-BiGJoin algorithms in Timely Dataflow and evaluate their performances extensively. Our evaluations include comparisons against an optimized single threaded algorithm, an existing shared-parallel system, and two existing distributed systems specialized for evaluating subgraph queries. We show that our approach can monitor complex sugraphs very efficiently on graphs with up to 64B edges on a cluster of 16 machines using just over eight bytes per edge.  Graphs at the scale we process are significantly larger than graphs used in previous work. 
We note that our algorithms can also be easily used in both existing distributed bulk synchronous parallel systems, such as MapReduce~\cite{dean:mr} and Spark~\cite{zaharia:spark}, as well as streaming systems, such as Storm~\cite{toshniwal:storm} and Apache Flink~\cite{carbone:flink}.  

We end this section with a note on our theoretical contributions.

\noindent {\bf Delta-GJ's Optimality (Theorem~\ref{thm:delta-gj}):} Our Delta-BiGJoin algorithm is a distributed version of a new incremental view maintenance algorithm we develop for join queries called {\em Delta-GJ}. We prove that under insertion only workloads Delta-GJ is worst-case optimal. When we distribute Delta-GJ in Delta-BiGJoin, we achieve worst-case optimality in terms of communication as well.
 
\noindent {\bf BiGJoin-S's Optimality (Theorem~\ref{thm:bigjoin-s}):} 
The challenge in the distributed setting is to achieve optimality in cumulative bounds while requiring low memory, e.g., $O(\frac{\IN}{w})$, and workload per worker.
Indeed, a naive ``distributed'' algorithm can send all of the input to one worker $w^*$ and use a sequential worst-case join algorithm. This algorithm would achieve all of the optimality guarantees we seek but without balancing the workload in the cluster. 
 
 BiGJoin achieves cumulative worst-case optimality and in our real-world data sets and queries achieves good work\-load-balance and low per-worker memory. However, on adversarial inputs it can lead to a single worker performing most of the work. We address this theoretical shortcoming with BiGJoin-S. Specifically, 
BiGJoin-S is the first distributed join algorithm that has worst-case communication and computation costs and achieves workload-balance across workers on every query. In addition, BiGJoin-S achieves these guarantees with as low as  $O(\frac{\IN}{w})$ memory per-worker.
In prior work, reference~\cite{koutris:worst-case-parallel} had shown that variants of the Shares algorithm have the same guarantees only for certain queries, e.g., cycles.
We provide a detailed comparison of BiGJoin-S with the algorithm in reference~\cite{koutris:worst-case-parallel} in 
\iftechreport
Appendix~\ref{app:wco-load}.
\else
our longer technical report~\cite{ammar:bigjoin-tr}.
\fi

\vspace{10pt}
\section{Preliminaries}
\label{sec:preliminaries}

\subsection{Notation}
\label{subsec:notation}

We present our algorithms in the general setting when they process general multiway equi-join queries, also referred to as {\em full conjunctive queries}. Let $Q$ be a query over $n$ relational tables, $R_1$, ..., $R_{n}$, where each $R_i$ is over a subset of $m$ attributes $a_1, \ldots, a_{m}$. We let $\IN = \Sigma_i |R_i|$ be the size of the input.
We write $Q$ as: 
$$Q(a_1, ..., a_{m}) := R_1(a_{11}, ..., a_{1r_1}),...,R_{n}(a_{n1}, ..., a_{nr_n})$$

\begin{figure}[t]
\begin{lstlisting}[mathescape]
$P_0$={}
for ($j$ = $1$... $m$):
   $P_j$={}
   for ($p \in P_{j-1}$):
       // $\cap$ $\text{below is performed starting from smallest}$Ê $Ext_j^i(p)$
       $ext_p = \cap Ext_j^i (p)$
       $P_j$ = $P_j \cup ext_p$ 
\end{lstlisting}
    \vspace{-5pt}
\caption{Pseudocode of GJ.}
 \label{fig:gj-pseudocode}
\end{figure}

\subsection{Generic Join}
\label{subsec:gj}
We base our work on the GJ algorithm (Figure~\ref{fig:gj-pseudocode}). Given a query $Q$, GJ consists of the following three high-level steps:
\squishlist
\item {\bf Global Attribute Ordering:} GJ first orders the attributes. Here we assume for simplicity the order is $a_1, \ldots, a_m$. We  will have a stronger preference on the order, but everything that follows remains correct if the attributes are arbitrarily ordered.

\item {\bf Extensions Indices:} Let a {\em prefix $j$-tuple} be any fixed values of the first $j < m$ attributes. For each $R_i$ and  $j$-tuple $p$ only some values for attribute $a_{j+1}$ exist in $R_i$. Let the {\em extension index} $Ext^i_j$ map each $j$-tuple $p$ to values of $a_{j+1}$ matching $p$ in $R_i$:
$$ Ext^i_j : (p=(a_{1}, .., a_{j})) \rightarrow \{ a_{j+1} \} \; . $$
Extension indices need three properties for the theoretical bounds of GJ: for a given $p$ we can retrieve (i) the size $|Ext^i_j(p)|$ in constant time, (ii) the contents of $Ext^i_j(p)$ in time linear in its size, and (iii) check that a value $e$ of attribute $a_{j+1}$ exists in $Ext^i_j(p)$ in constant time. Throughout the text we denote by $Ext^i_j(p \bullet e)$ the operation of checking of value $e$ in $Ext^i_j(p)$. These properties are satisfied by many indices, for example hash tables. 

\item {\bf Prefix Extension Stages:} GJ iteratively computes intermediate results $P_1\ldots P_m$, where $P_j$ is the result of $Q$ when each relation is restricted to the first $j$ attributes in the common global order. GJ starts from the singleton relation $P_0$ with no attributes, determines $P_{j+1}$ from $P_j$ using the extension indices, and ultimately arrives at $P_m = Q$. 
Specifically, for each {\em prefix} $j$-tuple $p \in P_j$, GJ determines the (possibly empty) set of $(j+1)$-tuples extending $p$ by intersecting the $Ext^i_j(p)$ {\em extension sets} of each relation $R_i$ containing $a_{j+1}$. This is done by proposing candidate extensions from the smallest of the sets, and then intersecting each candidate with the extension indices of the remaining relations. Starting from the smallest set, and in general performing this intersection in time proportional to the size of the smallest set ensures worst-case optimal run-time.

\end{list}

An example of GJ is given in 
\iftechreport
Appendix~\ref{app:gj-example}.
\else
our longer technical report~\cite{ammar:bigjoin-tr}.
\fi
 We next re-state a theorem from \cite{ngo:survey} using our notation:

\begin{theorem}~\cite{ngo:survey}
\label{thm:gj}
For any query $Q$ comprising relations $R_1 \ldots R_n$ and attributes $a_1 \ldots a_m$, and any ordering of attributes, if  $Ext^i_j$ indices satisfy the three properties discussed above, GJ runs in time $O(m n  MaxOut_Q)$.
\end{theorem}

\subsection{Massively Parallel Computation Model}
\label{subsec:mpc}
Massively Parallel Computation (MPC)~\cite{beame:communication, beame:skew, koutris:worst-case-parallel} is an abstract model of distributed bulk synchronous parallel systems. 
Brief\-ly there are $w$ workers in a cluster. The input data is assumed to be equally distributed among the workers arbitrarily. The computation is broken down into a series of {\em rounds}, where in each round the workers first perform some local computation and then send each other messages. 
The complexity of algorithms are measured in terms of three parameters: (1) {\em r}: the number of rounds; (2) {\em L}: the maximum {\em load} or messages any of the workers receives in any of the rounds; and (3) {\em C}: the total communication, i.e., sum of the loads across all rounds.

We extend  MPC with a fourth parameter $M$ that measures the memory that an algorithm uses. 
Let $LocM^{t}_k$ be the local memory that worker $k$ requires in round $t$, excluding the output tuples. In our setting, $LocM^{t}_k$ will be the load $L$ of worker $k$ in round $t$ and the amount of input data worker $k$ has indexed. $M$ is then the $\max_{t=1,...,r}\Sigma_{k=1,...,w}LocM^{t}_k$.
 We assume output tuples are written to a storage outside the cluster and do not stay in memories of workers. This is because any correct algorithm incurs this cost.

For simplicity, similar to prior work~\cite{afrati:mrmodel, beame:communication, koutris:worst-case-parallel} our unit of communication and memory will be tuples and prefixes, instead of bits, and we assume that tuples and prefixes have a common unit size.

\subsection{Timely Dataflow}
\label{subsec:timely}
Timely Dataflow~\cite{murray:naiad} is a distributed data-parallel dataflow system, in which one connects dataflow {\em operators} describing computation using dataflow edges describing communication. The operators are {\em data-parallel}, meaning that their input streams may be partitioned by a provided key, and their implementations may be distributed across multiple workers. All operators are distributed across all workers, and each worker is responsible for the execution of some fraction of each operator, which allows our algorithms to share indices (of the underlying relations) between operators.

Timely Dataflow is a dataflow system in the sense that computation occurs in response to the availability of data, rather than through centralized control.
The {\em timely} modifier corresponds to the extension of each operator with information about logical progress through the input streams, roughly corresponding to {\em punctuation} or {\em watermarks} in traditional stream processing systems. 
Importantly for the current paper,  operators can delay processing inputs with some timestamps until others have finished, which can be used to synchronize the workers and ensure that the work queues of downstream operators have drained, an important component of ensuring a bounded memory footprint.

\section{Algorithms}
\label{sec:bigjoin}

Our algorithms are based on a common dataflow primitive that extends prefixes $P_{j}$ to $P_{j+1}$. We first describe a naive version of the primitive that explains the overall structure (and is closest to the implementation we evaluate). We then develop the BiGJoin and Delta-BiGJoin algorithms using this core primitive. We then modify the primitive and develop BiGJoin-S to achieve workload-balance and skew-resilience.

\subsection{Dataflow Primitive}
\label{subsec:dataflowprimitive}

The core dataflow primitive starts from a collection of $P_j$ tuples stored across $w$ workers, and produces the $P_{j+1}$ tuples across the same workers. We first describe a dataflow that closely tracks the GJ algorithm, starting from the full collection $P_j$ and producing the full collection $P_{j+1}$. We will need to modify this dataflow in several ways to achieve both memory boundedness and workload balance across workers to achieve our theoretical bounds, but this simpler description is instructive and empirically useful.

\subsubsection{A synchronous implementation} 

We first describe the dataflow primitive as a sequence of steps, where workers 
execute each step to completion and synchronize between each step (corresponding to a round in BSP terms). Naive execution of these steps may produce very large amounts of data between steps and require very large memory in the workers.

\squishlist
\item {\em Initially}: The tuples of $P_j$ are distributed among the $w$ workers arbitrarily. Each prefix $p$ is transformed into a triple $(p, \infty, \bot)$ capturing the prefix, the currently smallest candidate set size, and the index of the relation with that number of candidates.

\item {\em Count minimization}: For each $R_i$ binding attribute $a_{j+1}$, in order:
Workers exchange the triples by the hash of $p$'s attributes bound by $R_i$, placing each triple at the worker with access to $Ext^i_j(p)$. Each worker, for each triple updates the smallest count and introduces its own index if $|Ext^i_j(p)|$ is smaller than the recorded smallest count. Each triple is then output as input of the count minimization for the next relation. In the end we have a collection of triples $(p, min{\text -}c, min\text{-}i)$ indicating for each prefix the relation with the fewest extensions.

\item {\em Candidate Proposal}: Each worker exchanges triples using a hash of $p$'s attributes bound by $R_{min\text{-}i}$. Each worker now produces for each triple $(p, min\text{-}c, min\text{-}i)$ it has, and each extension $e$ of $p$ in $Ext^{min\text{-}i}_j(p)$, a candidate $(j+1)$-tuple $(p \bullet e)$.

\item {\em Intersection}: For each relation $R_i$ binding attribute $a_{j+1}$, in order: Workers exchange the candidate $(p \bullet e)$ tuples by the hash of $(p \bullet e)$'s attributes bound by $R_i$. Each worker consults $Ext^i_{j}(p)$. If $e$ exists $(p \bullet e)$ is produced as output otherwise it is discarded.

\end{list}
Figure~\ref{fig:dataflow-primitive} shows the operators of this dataflow primitive. In the figure, the vertical lines annotated with $s$ indicate synchronization points. 
These steps, executed in sequence would be a synchronous BSP implementation of the computation extending $P_j$ to $P_{j+1}$, which we could repeat until we arrive at $P_m = Q$.
The random access working set of these operators are only the extension indices; all inputs and outputs are processed sequentially. Nonetheless, the sizes of the inputs and intermediate outputs to operators could be quite large requiring large memory/storage, which we address next.

\subsubsection{A batching optimization to reduce memory}
\label{subsub:batching}
Notice that the {\code Proposal} operator is the only operator that may produce more output than it consumes as input and increase the memory usage of the system. We can fix this with a simple batching optimization. Instead of producing all of the proposals for each $P_j$ prefix they have, each {\code Proposal} operator produces its candidate extensions in batches of $B'$. 
The remaining extensions are produced in the subsequent invocations. This may leave some prefixes only partially extended. To keep track of these partial extensions,
we store $(p, min\text{-}c, min\text{-}i, rem{\text-}ext)$ quadruples where $rem{\text-}ext$ is the {\em remaining extensions} metadata. Letting $B=wB'$, this ensures that the dataflow has at most $B$ queued elements at any time across the workers, as the $B$ proposals created by {\code Proposal} operators are retired before any more are produced. The {\code Count} and {\code Intersect} steps remain unchanged. 

\subsubsection{A streaming implementation}
Except for the {\code Proposal} operators, the operators  described above do not need to synchronize. Specifically, instead of synchronizing, the {\code Count} and {\code Intersect} operators can produce outputs as inputs to their next operators as soon as they receive inputs. This leads to a streaming implementation, which can improve performance in practice. 
When implementing the above batching optimization however, the {\code Proposal} operators need to synchronize and be notified that they can produce another batch of extensions. 

\begin{figure}[t!]
  \centering
	\includegraphics[width=0.5\textwidth]{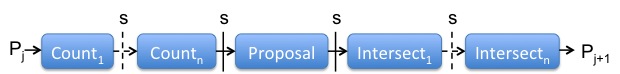}
  \vspace{-15pt}
  \caption{Dataflow Primitive.}
  \label{fig:dataflow-primitive}
  \vspace{-15pt}
\end{figure}

\subsection{Joins on Static Relations: BiGJoin}
\label{subsec:bigjoin}

We now describe how to use our dataflow primitive to build a dataflow for evaluating queries on static graphs. First, we order the attributes arbitrarily, and build indices over each relation for each prefix of its attributes in the global order. Next, we assemble the dataflows for extending each $P_j$ to $P_{j+1}$ for each attribute $a_j$, so that starting from an empty input tuple $()$ we produce streams of prefixes $P_j$, out to $P_m = Q$. Finally, we introduce the empty tuple to start the computation, producing the stream of records from $Q$ as output. We use the batching optimization described above and when deciding which batch of $P_j$ to $P_{j+1}$ extensions to invoke next, we pick the largest $j$ value such that at least one worker has $B'$ prefixes to propose. We refer to this algorithm as BiGJoin. The next lemma summarizes the costs of BiGJoin:

\begin{lemma}
\label{obs:bigjoin}
Given a query $Q$ over $m$ attributes and $n$ relations, the communication and computation cost of BiGJoin equals that of computation of GJ and is $O(mnMaxOut_Q)$. Let $B'$ be a batching parameter and let $B=wB'$. The cumulative memory BiGJoin requires is $O(m\IN + mB)$, and the number of rounds of computation BiGJoin takes is $O(\frac{mnMaxOut_Q}{B'})$.
\end{lemma}
\vspace{-3pt}
The proof of this lemma, presented in 
\iftechreport
Appendix~\ref{app:thm-bigjoin}
\else
our longer technical report~\cite{ammar:bigjoin-tr}
\fi
, is based on the fact that each operation that BiGJoin does on each tuple corresponds to an operation in the serial execution of GJ and small enough batches can keep the memory footprint very low. In essence, BiGJoin inherits its computation and communication optimality from GJ. Moreover, as we will demonstrate in Section~\ref{sec:evaluation}, in practice BiGJoin also achieves good workload-balance across the workers in the cluster. However, on adversarial inputs BiGJoin cannot guarantee workload balance.  
We will address this theoretical shortcoming in Section~\ref{subsec:bigjoin-s} to achieve one of our main theoretical results.

\subsection{Joins on Dynamic Relations: Delta-BiGJoin}
\label{subsec:delta-bigjoin}
We next show how to use our dataflow primitive to maintain join queries over dynamic relations, which we use to maintain subgraph queries on dynamic graphs. We first describe a new incremental view maintenance (IVM) algorithm for join queries called {\em Delta-GJ} and then describe its distributed version Delta-BiGJoin.

\subsubsection{Delta-GJ}
\label{subsubsec:deltagj}
Let $Q$ be a query and consider a setting where for each relation $R_i$ we have a change $\Delta R_i$, corresponding to the addition and deletion of some records in $R_i$. We begin by reviewing an incremental view maintenance technique based on delta queries from references~\cite{blakeley:ivm, gupta:ivm}. 
Let's assume that tuples in each $\Delta R_i$ are labeled such that we can tell the inserted tuples apart from the deleted ones. Let $R_i'$ be $R_i + \Delta R_i$, where the union operation removes a tuple $t$ in $R_i$ if $\Delta R_i$ contains a deletion of $t$. Let $Out$ and $Out'$ be the output of $Q$ before and after the updates, respectively. Then consider the following $n$ delta queries:
\begin{align*}
&dQ_1 := \Delta R_1, R_2, R_3, ..., R_n \\
&dQ_2 := R_1', \Delta R_2, R_3, ..., R_n \\
&dQ_3 := R_1', R_2', \Delta R_3, ..., R_n \\
&\phantom{xxxxxxxxxxx}... \\ 
&dQ_n := R_1', R_2', R_3', ..., \Delta R_n
\end{align*}
We assume output tuples that emerge from inserted and deleted tuples are labeled as inserted and deleted, respectively. It can be shown that the union of the $n$ queries above are exactly the changes to the output of $Q$, i.e., $Q' \setminus Q = dQ_1 + dQ_2 + ... + dQ_n$~\cite{blakeley:ivm, gupta:ivm}.

Delta-GJ runs the $n$ delta queries indepenpendently, where each $dQ_i$
is executed using GJ. Note that Delta-GJ's correctness, i.e., that it finds the correct
differences to the output of $Q$, simply follows from the correctness of the delta query
technique~\cite{blakeley:ivm, gupta:ivm}. However, in order to prove that Delta-GJ is efficient,
we need to order the attributes of each $dQ_i$ in a specific order. Specifically, for $dQ_i$, Delta-GJ picks an attribute ordering that starts
with any permutation of $R_i$'s attributes $a_{i1}, a_{i2}, ..., a_{ir_i}$
and an arbitrary order for the remaining $m$$-$$r_i$ attributes. 
The next theorem states that under insertion-only workloads, Delta-GJ is a worst-case optimal IVM algorithm for join queries. The proof is given in 
\iftechreport
Appendix~\ref{app:thm-deltagj}.
\else
our longer technical report~\cite{ammar:bigjoin-tr}.
\fi

\begin{theorem}
\label{thm:delta-gj}
Consider a query $Q$ and a series of $z$ updates that only consist of
inserting tuples to the input relations of $Q$.  Let $R_i(z)$ denote the relation $R_i$ after the $z$'th update. 
Then the total computation cost of Delta-GJ
is $O(mn^2MaxOut_Q)$, where $MaxOut_Q$ is the AGM bound of $Q$ on $R_i(z)$.
\end{theorem}
It is harder to characterize the performance of Delta-GJ under workloads with both insertions and deletions because ``problematic" rec-ords might require a lot of work and could simply be repeatedly added and removed. 
A more precise characterization of Delta-GJ under arbitrary workloads is left as future work.
We note that an incremental version of the Leapfrog TrieJoin algorithm~\cite{veldhuizen:inc-lftj} also achieves worst-case optimality under insertion-only workloads but by maintaining indices that can be super-linear in the size of the inputs. Delta-GJ's indices are linear in the size of the inputs.\footnote{In a separate paper, one of the co-authors and his colleagues have used Delta-GJ to support triggers in the context of an active graph database called Graphflow~\cite{kankanamge:graphflow}. The Graphflow paper cited a previous partial technical report version of the current paper.}

\subsubsection{Delta-BiGJoin}
We next describe how we parallelize Delta-GJ in the distributed setting.
We have a separate dataflow for each $dQi$ that is a $dQ_i$-specific variation of the BiGJoin dataflow from Section~\ref{subsec:bigjoin}.
By ordering the attributes of $dQ_i$ starting with the attributes of $R_i$, we can seed the computation with the elements of $\Delta R_i$, instead of $()$, which is expected to be much smaller than the other relations in $dQ_i$. 
Importantly, we only need to {\em maintain} the indices as changes occur, rather than fully rebuilding them. The resulting cost is proportional to the number of changes (for rebuilding indices) and the number of prefixes in the delta queries as we evaluate them. 

The next lemma is proved in 
\iftechreport
Appendix~\ref{app:thm-deltabigjoin}. 
\else
our longer technical report~\cite{ammar:bigjoin-tr}. 
\fi

\begin{lemma}
\label{obs:delta-bigjoin}
Consider a series of $z$ insertion-only updates to the input relations of a query $Q$. Let $R_i(z)$ denote the relation $R_i$ after the $z$'th update and $\IN(z)$ be $\sum_i |R_i(z)|$. Then, given a batch size $B'$ and letting $B=wB'$, Delta-BiGJoin's communication and computation cost is $O(mn^2MaxOut_Q)$. The cumulative memory Delta-BiGJoin uses is $O(mn\IN(z) + mB)$. In MPC terms, the number of rounds of computation Delta-BiGJoin takes is $O(\frac{mn^2MaxOut_Q}{B'} + zmn^2)$.
\end{lemma}

\subsection{A Work-balanced Dataflow: BiGJoin-S}
\label{subsec:bigjoin-s}
As we demonstrate in Section~\ref{sec:evaluation}, BiGJoin and Delta-BiGJoin perform very well on the real-world queries and datasets we experimented with. However they have an important theoretical shortcoming. Specifically, they do not guarantee that the workloads of the workers are balanced. Indeed, it is easy to construct skewed inputs where most of the work could even be performed by a single worker.
We next modify our dataflow primitive to ensure workload balance across workers. We note that the contributions of this section are theoretical. An implementation and evaluation of these techniques are left for future work.

There are three sources of imbalance in our dataflow primitive:
\squishlist
\item[1.] {\em Sizes of extension indices:}  
Recall that $Ext_j^i$ are distributed randomly. Yet for each prefix $p$, a single worker stores the entire $Ext_j^i(p)$ (the $a_{j+1}$ extensions of $p$). In graph terms, this corresponds to a single worker storing the entire adjacency list of a vertex. On skewed inputs, this may generate imbalances in the amount of data indexed at each worker.

\item[2.] {\em Number of Proposals:} After count minimization, each worker gets a set of $(p, min\text{-}c, min\text{-}i, rem\text{-}ext=min\text{-}c)$ quadruples where $p$ is  a $P_j$ prefix to extend. Even if each worker has to extend the same number of prefixes, each worker might have to do imbalanced amount of proposals of $(p \bullet e)$ candidate extensions because the counts might be very different. 

\item[3.] {\em Number of Index Lookups:} When minimizing the counts of a $P_j$ prefix $p$, producing the candidate proposals, or  intersecting the $(p \bullet e)$ candidate extensions with $Ext^i_j$, prefixes and candidate extensions are
routed to the worker that holds $Ext^i_j(p)$ based on the hash of $p$'s attributes that are bound by $R_i$. If there are many prefixes whose attribute values that are bound by $R_i$ are the same, there may be an imbalance in the number of prefixes and extensions each worker receives. For example, consider a triangle query where all triangles involve some specific vertex $v^*$, then every $P_2$ prefixes could be routed to a single worker to access $v^*$'s count.

\end{list}
We show how to fix these sources of imbalance without asymptotically affecting the other costs of BiGJoin.

\subsubsection{Skew-resilient Extension Indices}
We distribute the contents of $Ext_j^i(p)$ across workers, instead of storing at a single worker. Specifically, we store three indices.  

\squishlist
\item $C_j^i(p)$ {\em Count Index}: Stores the size of $Ext_j^i(p)$. This index is distributed randomly by the hash of prefixes $p$.
\item$Ext\text{-}Res_j^i((p, k))$ {\em Extension Resolver Index}: 
Let $\{e_1, ..., e_c\}$ be the $a_{j+1}$ extensions of $p$ in $R_i$. 
We use $Ext\text{-}Res_j^i((p, k))$, for $k=1\ldots c$, to {\em resolve} the $k$th extension, i.e.,  $Ext\text{-}Res_j^i((p, k))$ $=e_k$. This index is distributed randomly by the hash of $(p, k)$ tuples. Essentially we distribute each element of $Ext_j^i(p)$ randomly across the workers.
\item $Ext_j^i((p \bullet e))$: As in the original extension indices, this index is used to lookup the existence of a particular extension $e$ of $p$ and is distributed randomly by the hash of $(p, e)$.
\end{list}
Since the contents of $Ext_j^i(p)$ are now randomly distributed across workers, these indices fix the first source of imbalance from above.

\begin{figure}[t!]
  \centering
	\includegraphics[width=0.5\textwidth]{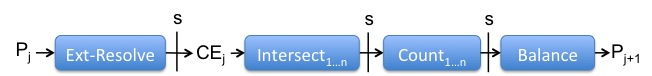}
  \vspace{-10pt}
  \caption{Work-balanced Dataflow Primitive.}
  \label{fig:balanced-dataflow}
  \vspace{-15pt}
\end{figure}

\iftechreport
\newpage
\fi

\subsubsection{Balance and Extension Resolving Operators}
We modify the dataflow primitive as shown in Figure~\ref{fig:balanced-dataflow}. 
Compared to BiGJoin's dataflow primitive, BiGJoin-S's dataflow primitive contains modifications of the {\code Count} and {\code Intersect} operators, does not contain the {\code Proposal} operator, and contains two new operators  {\code Balance} and {\code Extension-Resolve}. In Figure~\ref{fig:balanced-dataflow}, the $P_j$ tuples are $(p, min{\text-}i, start,  end)$ quadruples which indicate a range, indicated by $start$ and $end$ of candidate extensions that the worker holding the tuple should make for the prefix $j$-tuple $p$. The dataflow however takes $(p, k)$ tuples where $k \in [start, end]$ and produces a set of $(p', min{\text-}i, start,  end)$ where $p' \in P_{j+1}$. The $(p, k)$ tuples move along the operators as follows:

\squishlist

\item {\code Extension-Resolve:} Each worker for each $(p, $ $min{\text-}i, \\start,$ $end)$ tuple and $k \in [start, end]$, {\em resolves} $(p, k)$ by consulting the $Ext\text{-}Res_{j}^{min{\text-}i}$ indices and gets back a candidate extension $(p \bullet e)$. As in BiGJoin this happens in batches of $B'$ $(p, k)$ tuples per worker. Recall that the lookup for $(p, k)$ is made to the worker that holds $Ext\text{-}Res_{j}^{min{\text-}i}$ based on the attributes of $p$ bound by $R_{min{\text-}i}$. Although each $(p, k)$ is distinct, there may be skew in the attribute values of $(p, k)$ that are bound by $R_{min{\text-}i}$. To guard against this, workers first locally aggregate the $Ext\text{-}Res_{j}^{min{\text-}i}$ requests they will make to the same relation ${R_{min{\text-}i}}$ with the same lookup key (the attributes of $p$ bound by $R_{min{\text-}i}$) and $k$ value, and send only one request instead. 
This and a similar aggregation optimization in the {\code Intersect} and  {\code Count} operators fix the third source of imbalance from above. 
Upon receiving the answers to their requests, workers have the set of candidate extensions, which we refer to as $CE_{j}$. 

\item {\code Intersect:} Instead of routing each $(p\bullet e)$ through the $Ext_j^i$ indices one by one as done by BiGJoin's {\code Intersect} operator, each worker {\em manages} each $(p\bullet e)$ candidate extension it initially holds. Specifically for each $R_i$ that contains $a_{j+1}$, in synchronous rounds, each worker does a distributed lookup of $(p\bullet e)$ in $Ext_j^i$ by sending $(p\bullet e)$ to the worker that holds  $Ext_j^i(p \bullet e)$ and gets the tuple back with a yes/no label. Similar to the {\code Extension-Resolve} operator, workers locally aggregate the requests they will make with the same key.

Recall that the worker that holds $Ext_j^i(p \bullet e)$ is based on the hash of the attributes of $(p \bullet e)$ bound by $R_i$ (possibly including the value $e$). Although each $(p \bullet e)$ candidate extension is distinct, there may be skew in these projections. 
To guard against this, before sending their lookup requests, workers first aggregate the projections of all of their $CE_j$ candidate extensions and for each possible projection sends at most one lookup request.  
After at most $n$ intersections, each $(p \bullet e)$ either becomes a $P_{j+1}$ prefix or is discarded if it does not successfully intersect an $R_i$.

\item {\code Count:} For each $P_{j+1}$ prefix, workers compute the $(p'=(p\bullet e), min\text{-}c,$ $min\text{-}i)$ triples, after at most $n$ synchronous rounds, by looking up $p'$ in the $C^i_{j+1}$ indices by aggregating lookups with the same key. Similar to the above {\code Intersect} operator and unlike BiGJoin's {\code Count} operator, instead of routing the prefixes through each $C^i_{j+1}$ index, the workers manage the triples. 

\item {\code Balance:} For each $(p', min\text{-}c, min\text{-}i)$ tuple, there is {\code min-c} number of proposals and following intersections to make. There may be an imbalance in how much intersection work each worker gets after the count minimization. To balance this intersection work, each worker deterministically distributes its total proposal work among the other workers. Each worker $w_{\ell}$ first finds the target intersection work amount $T$ to distribute and gives $T/w$ proposal and intersection work (with a +/- 1 difference) to each other worker $w_{\ell'}$. This is done by sending $(p', min{\text-}i, start,$  $end)$ tuples to $w_{\ell'}$. The $start \le end$ $\le min{\text-}c$ indicate the range of extensions among the $min{\text-}c$ total candidate extensions of $p'$ that the receiving worker $w_{\ell'}$ is responsible for. At this point each worker gets a set of $(p', min{\text-}i, start,$  $end) \in P_{j+1}$ tuples. 
This fixes the second form of skew discussed above.

\end{list}
Similar to BiGJoin, we assemble this workload-balanced dataflow for extending $P_j$ to $P_{j+1}$ for each attribute $a_j$. 
We call this algorithm with batching optimization in the {\code Ext-Resolve} operator BiGJoin-S. When  deciding which batch of $P_j$ to $P_{j+1}$ candidate extensions to compute next, BiGJoin-S picks the largest $j$ value such that all of the workers have $B'$ candidate extensions to resolve and propose (instead of at least one as in BiGJoin).
The following theorem states that BiGJoin-S achieves 
workload balance over large enough, but logarithmic size, batch sizes, while asymptotically maintaining the optimality bounds of BiGJoin on any query and arbitrary datasets (so under any amount of skew in inputs). It is the first algorithm to achieve these bounds for arbitrary queries and datasets. A detailed comparison of its costs against a variant of the Shares algorithm is provided in
\iftechreport
Appendix~\ref{app:wco-load}. 
\else
our longer technical report~\cite{ammar:bigjoin-tr}.
\fi
The proof is technical and provided 
\iftechreport
in Appendix~\ref{app:thm-bigjoin-s}.
\else
also in our longer technical report~\cite{ammar:bigjoin-tr}.
\fi

\begin{theorem}
\label{thm:bigjoin-s}
Suppose $\frac{B'}{w} \ge \max\{w, \log(\IN \times MaxOut_Q)\}$ and let $B=wB'$. Then BiGJoin-S has the following costs:
\squishlist
\item Cumulative computation and communication cost of \linebreak $O(mnMaxOut_Q)$ and memory cost of $O(mn\IN + mB)$.
\item $O(\frac{mnMaxOut_Q}{B})$ rounds of computation.
\item With at least  probability $1 - O(\frac{1}{\IN})$, each worker performs $O(B')$ communication and computation in each round of the algorithm. In MPC terms, the load of BiGJoin-S is $O(\frac{\mathrm{mnIN}}{w} + mB')$, so assuming $B' < \frac{IN}{w}$, BiGJoin-S has optimal load.
\end{list}
\end{theorem}
We note that we can make Delta-BiGJoin also skew-resilient under large enough updates by using BiGJoin-S with delta queries instead of BiGJoin. We need the update sizes, i.e., the size of the $\Delta R_i$, to be large enough to make Delta-BiGJoin-S skew-resilient. For example if each update to the relations contain a single tuple, then the amount of work to maintain the query results could be too small to possibly distribute equally across workers.

\section{Implementation}
\label{sec:timely-impl}

In this section we describe our implementations of BiGJoin and Delta-BiGJoin in Timely Dataflow. Although our implementations are tailored for evaluating subgraph queries, so the input relations are binary relations consisting of the edges of an input graph, the underlying machinery nonetheless is suitable for more general quer\-ies. 
We will demonstrate this in Section~\ref{subsec:seed} when using our algorithm to take as input a ternary relation.
We start by developing the prefix extension dataflow primitive as a Timely fragment. Our implementations can be found here~\cite{timely-bigjoin}. 

\subsection{Prefix Extension in Timely Dataflow}

Our approach to prefix extension follows the primitive from Section~\ref{subsec:dataflowprimitive}: we will assemble a dataflow fragment that starts from a stream of prefixes of some number $j$ of attributes, and produces as output the corresponding stream of prefixes resulting from the extension of each input prefix by the relations constraining the attribute $a_{j+1}$ in terms of the first $j$ attributes. As we explain in Section~\ref{subsec:delta-bigjoin-t}, for our Delta-BiGJoin implementation, the prefixes are tagged with a timestamp and a signed integer, reflecting the time of change and whether it is an addition or deletion, respectively.

Prefix extension happens through three methods acting on strea-ms, corresponding to the three steps described in Section~\ref{subsec:dataflowprimitive}: count minimization, candidate proposal, and intersection. Each of these steps is implemented as a sequence of operators, each of which corresponds to one of the relations constraining attribute $a_{j+1}$ . Each operator will consult some indexed form of the relation it represents, and requires the prefixes in its input stream to be shuffled by the corresponding attribute, so that prefixes arrive at the worker that store the appropriate fragment of the index. Importantly, we use the same partitioning for each relation and attribute in that relation, so that any number of uses of the relation in the query require only one physical instance of each index. In the case of graph processing, this means we keep only a forward and reverse index, storing respectively the outgoing and incoming neighbors of each vertex.

\subsubsection{Count Minimization} The implementation of this step is straightforward and follows our description in Section~\ref{subsec:dataflowprimitive} directly. There is a sequence of operators and each one represents one relation $R_i(a_{j+1}, a_k)$ or $R_i(a_k, a_{j+1})$, where $k \le j$. The operator takes $(p, c, i)$ triples as input. Let $v^*=\Pi_{a_k}p$. The operator updates the count $c$ if the size of $v^*$'s outgoing neighbors (if $R_i=R_i(a_{j+1}, a_k)$), or incoming neighbors (if $R_i=R_i(a_k, a_{j+1})$), is less than $c$. At the end we identify the $(p, min{\text-}c, min{\text-}i)$ triples but then send only $p$ to a stream for $R_i$ (explained next).

\subsubsection{Candidate Proposal} This step is implemented by a single operator that divides its stream of input prefixes into one stream for each relation $R_i$, by the $min{\text-}i$ index identified in the previous stage. Suppose $R_i=R_i(a_{j+1}, a_k)$. Then an input $p$ whose $min{\text-}i$ was $i$, where $v^*=\Pi_{a_k}p$, will be part of the stream for $R_i$ and be extended to a tuple $(p \bullet \{e_1, ..., e_c\})$ containing the {\em set of candidate extensions}, which are $v^*$'s outgoing neighbors. When $R_i=R_i(a_k, a_{j+1})$, we use the incoming neighbors of $v^*$ instead. This deviates from our description where we had flattened this tuple for simplicity of explanation and had $c$ separate $(p \bullet e)$ candidate extensions. These extensions are sent through a single output stream for the next stage.

\subsubsection{Intersection} The stream of pairs of prefix and candidate extensions go through a sequence of operators, one for each involved relation $R_{i}$, each of which intersects the set of candidate extensions with an appropriate neighbor list of a vertex and removes those extensions that do not intersect.
The result is a stream of pairs of prefix and valid extensions, successfully intersected by all relations. The extensions are flattened to a list of prefixes for the next stage except if they are the final outputs, they are output in their compact representation.

\subsection{The BiGJoin Dataflow}

The dataflow for enumerating subgraphs in a static graph applies a sequence of prefix extension stages, each corresponding to an attribute in the global attribute order. For simplicity, we fix the global order so that the first two attributes are connected by an edge, which allows us to seed the stream of prefixes with length-two prefixes read from the edges themselves. This is equivalent to starting the extensions from $P_2$ instead of the empty tuple $() \in P_0$. All other attributes are extended using the prefix extension dataflow fragment we described above.

The indices used by the workers are static, and we simply memo-ry-map in a pre-built index. For simplicity we use the whole graph, which means we can easily vary the number of workers without changing the index used. One could alternately partition the graph and provide each worker with its own index, but the graphs on which we evaluate the static computations are rather small. For larger graphs, such as those we consider with DeltaBiGJoin, we build the indices as part of the computation, distributing the data to only the workers that require it.

The execution of the BiGJoin dataflow happens in batches, where we feed some number of prefixes into the dataflow and await their results before introducing more prefixes. This batching allows some control over the peak memory requirements, but does not guarantee that in the course of processing a batch we do not produce unboundedly many intermediate results. To manage back-pressure more precisely one can use the batching  techniques described in~\cite{lattuada:faucet}, which allow the {\em Proposal} stages to wait until the downstream data\-flow has drained as our batching optimization from Section~\ref{subsub:batching} does, but we have not implemented them for our evaluation as our basic input batching worked well in our evaluations.

\subsection{The Delta-BiGJoin Dataflow}
\label{subsec:delta-bigjoin-t}
The dataflow for finding subgraphs in a dynamically changing graph is more complex than for a static graph, along a few dimensions. First, as described in Section~\ref{subsec:delta-bigjoin}, we will have an independent dataflow for each $dQ_i$. Each dataflow is responsible for changes to each logical relation $R_i$ in the query, i.e., one for each of the edges in the subgraph query. Second, although these dataflows may execute concurrently, we will logically sequence them so that each dataflow computes the delta query as if executed in sequence (to resolve simultaneous updates correctly). Third, our index implementation will be more complicated, as it must support changes as well as the multi-versioned interaction required by the logical sequencing above.

We have a dataflow for each $dQ_i$, each of which uses a different global attribute order as described in Section~\ref{subsec:delta-bigjoin}. Although there are several dataflows with different attribute orders, each operator only requires access to either the {\em forward} or {\em reverse} edge index.

Each delta query dataflow $dQ_i$ computes changes in the outputs made to relation $R_i$ with respect to the other relations. Recall that $dQ_i$ uses the ``new" versions $R_i' = R_i + \Delta R_i$ for $i < j$ and the ``old" versions $R_i$ for $i > j$. This has the effect of logically sequencing the update rules, so that they are correct even if there are simultaneous updates to the input relations, something we expect in graph queries where the single underlying edges relation is re-used often. This use of new and old versions of the same index requires our implementation to be multi-versioned, if we want to only have a single copy of each index.

Our index implementation is a multi-version index, which tracks the accumulation of $(src, dst)$ pairs at various times and with various integer weights. It can respond to queries about the outgoing and incoming neighbors for a given key $v$ using updates at a target time. The updates are ``committed" when all tuples in the system have a timestamp greater than it (meaning the update will participate in all future accumulations for $v$); this information comes from Timely Dataflow's progress tracking infrastructure. 
The index maintains data in three regions: (i) a compacted index of committed updates, (ii) an uncompacted index of committed updates, and (iii) an ordered list of uncommitted updates. Committed updates are moved to the uncompacted index, which uses a log-structured merge list for each $v$, which can be compacted on a per-vertex basis to ensure that the amortized work for each vertex lies within the bounds prescribed for the worst-case optimality result. The compact index is formed from initial data during loading and in principle could be periodically re-formed by merging the uncompacted committed data. Practically, on large datasets we were not able to apply enough updates to make such re-compaction worthwhile, within reasonable experimentation timeframes.

The execution of the Delta-BiGJoin dataflow proceeds with the stream of batches of updates to the graph supplied as an input. Each of the tuples moving through a delta query dataflow has both a logical timestamp and a signed integer weight. The former allows us to work with multiple logical times concurrently, and to remain clear on which version of an index the prefix should be matched against. The integer weight allows us to represent both additions and deletions from the underlying relations.

\section{Evaluation}
\label{sec:evaluation}
We next evaluate the performance of our Timely Dataflow implementations of our algorithms on a variety of subgraph queries and large-scale static and dynamic input graphs. 

We first evaluate a reference computation (triangle finding) on several standard graphs using a few different systems, to establish a baseline for running time (Section~\ref{subsec:baseline}). With each system we quickly discover limits on their capacity; they struggle to load graphs at the larger end of the spectrum.  
We then study the scaling of our implementation as we vary the number of Timely workers both within a single machine as well as across multiple machines on a 64 billion-edge graph (Section~\ref{subsec:capacity-scaling}). We next demonstrate that several  optimizations that have been introduced in prior work can also be integrated into our algorithms to improve our algorithms (Section~\ref{subsec:seed}). Here we also show an experiment in which we use a ternary relation as input, demonstrating our algorithms' application to general relational queries. Finally, we study the effects of our batch size on performance and memory usage (Section~\ref{subsec:batch}).

We used both BiGJoin and Delta-BiGJoin in our experiments and refer to their Timely Dataflow implementations as BiGJoinT and Delta-BiGJoinT, respectively. Unless specified explicitly, we use a batch size of $100,000$ in all our experiments.

\subsection{Experimental Setup}
\label{subsec:exp-setup}

Table~\ref{table:data_sets} reports statistics of the graphs we use for evaluation. The sizes range from the relatively small but popular LiveJournal graph, with 68 million edges, up three orders of magnitude to the relatively large Common Crawl graph, with 64 billion edges. 
The abbreviations we use for the datasets are given in parentheses in Table~\ref{table:data_sets}. We used five queries: 
\squishlist
\item \texttt{triangle}:= e(a$_1$,a$_2$),e(a$_1$,a$_3$),e(a$_2$,a$_3$)
\item  \texttt{4-clique}:= e(a$_1$,a$_2$),e(a$_1$,a$_3$),e(a$_1$,a$_4$),e(a$_2$,a$_3$),e(a$_2$,a$_4$),e(a$_3$,a$_4$)
\item \texttt{diamond}:= e(a$_1$,a$_2$), e(a$_2$,a$_3$),e(a$_4$,a$_1$), e(a$_4$,a$_3$)
\item \texttt{house}:=$\text{ }$e(a$_1$,a$_2$),e(a$_1$,a$_3$),e(a$_1$,a$_4$),e(a$_2$,a$_3$),e(a$_2$,a$_4$),e(a$_3$,a$_4$),\linebreak e(a$_2$,a$_5$),e(a$_3$,a$_5$)\footnote{This is query $q6$ from the SEED reference~\cite{lai:seed} and is a 5-clique with two missing edges from one node.}
\item \texttt{5-clique}:= e(a$_1$,a$_2$),e(a$_1$,a$_3$),e(a$_1$,a$_4$),e(a$_1$,a$_5$),e(a$_2$,a$_3$),e(a$_2$,a$_4$),\linebreak e(a$_2$,a$_5$),e(a$_3$,a$_4$),e(a$_3$,a$_5$),e(a$_4$,a$_5$)
\end{list}

We note that the Common Crawl dataset has prohibitively large number of instances of each query. For example we estimate that there are more than $2.36 \times 10^{16}$ diamonds in Common Crawl, and enumerating all of them explicitly would take a prohibitively long time for any correct system. Instead, for the Common Crawl graph we focus on the \emph{incremental maintenance} of these queries, which can fortunately be performed without the initial computation of all answers.

For all experiments except one we used a local cluster of up to 16 machines. All machines have 2x Intel E5-2670 @2.6GHz CPU with 16 physical cores in total. Most machines have 256 GB memory, but we occasionally used a machine with 512 GB memory to accommodate single-machine experiments. Each machine has 10 Gigabit network interface. For experiments using EmptyHeaded (see Section~\ref{subsec:eh}), we used an AWS machine similar to our cluster machines (r3.8xlarge) and another machine with 1TB memory (x1.16xlarge) to accommodate EmptyHeaded's memory requirements when running the triangle query on the TW graph.

In all of our experiments we use one CPU core for each Timely worker. For each experiment we explicitly state how the workers are located, i.e., within a single machine, across machines, or both. 

\begin{table}
\centering
\begin{tabular}{| l | c | c | }
\hline
{\bf Name} & {\bf Vertices} & {\bf Edges}  \\
\hline
\cline{1-3}\cline{1-3}\cline{1-3}\cline{1-3}\cline{1-3}
LiveJournal (LJ)~\cite{snapnets}  & 4.8M & 68.9M \\
\cline{1-3}
Twitter (TW)~\cite{law} & 42M & 1.5B \\
\cline{1-3}
UK-2007 (UK)~\cite{law} & 106M & 3.7B \\
\cline{1-3}
Common Crawl (CC)~\cite{WDC} & 1.7B & 64B \\
\cline{1-3}
\end{tabular}
\caption{Graph datasets used in our experiments.}\label{table:data_sets}
\end{table}

\subsection{Baseline measurements}
\label{subsec:baseline}
We start with measurements of several existing approaches for finding subgraphs in static graphs. Our goal is to assess whether our implementations have relatively good absolute performance when evaluating queries in static graphs. We consider three baselines: (1) a single threaded implementation; (2) the shared-memory parallel EmptyHeaded system; and (3) the distributed Arabesque system. All of these implementations operate only on static graphs. None of these implementations are capable of working with our largest graph, and not all of them can evaluate our smaller graphs either. 

\subsubsection{COST}
COST~\cite{mcSherry:cost} (configuration that outperforms a single thread) is a metric to evaluate the parallelization overheads of a parallel algorithm or system. Specifically, COST of a parallel algorithm $A$ solving a problem $P$ is the number of cores that the algorithm needs to outperform an optimized single-threaded algorithm solving $P$. A small COST indicates that the system itself introduces little overhead, and the benefits of scaling are immediately realized.

\begin{figure}[!t]
\centering
\includegraphics[width=0.4\textwidth]{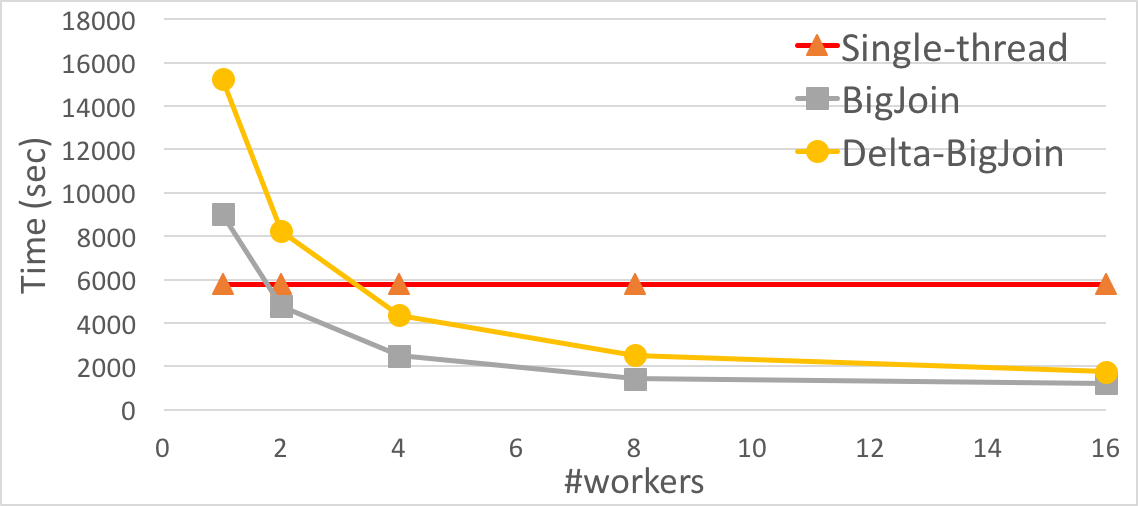}
\caption{BiGJoin and Delta-BiGJoin counting triangles in the Twitter graph, plotted with the time it takes our single-threaded implementation. Both approaches outperform the single-threaded implementation with small number of cores, and continue to improve from there. The Delta-BiGJoin performance lags slightly behind, as it uses more complex data structures to support updates.}
\label{fig:cost}
\end{figure}

In order to measure the COST of our algorithms, we implemented an optimized single-threaded triangle enumeration algorithm, that is based on GJ in Rust~\cite{rust}. We considered using the popular SNAP library~\cite{snapnets}, but found that our own single-threaded implementation was faster. We used the TW data set.  Figure~\ref{fig:cost} shows the optimized single-threaded, BiGJoinT, and Delta-BiGJoin-T implementations for the triangle query. Delta-BiGJoinT can find all triangles in a static graph by loading each of the edges as updates to the initially empty graph. However, it is expected to be slower than an algorithm that loads the whole graph first and then finds triangles. As seen there, the COST of our two implementations are 2 and 4 cores, respectively.

\subsubsection{EmptyHeaded}
\label{subsec:eh}

EmptyHeaded (EH) is a highly-optimized shared-memory parallel system evaluating subgraph queries on static graphs using GJ. EH evaluates queries using a mixture of GJ and binary join plans. The EH optimizer considers {\em generalized hypertree decompositions} of the query, which join multiple subsets of the relations using GJ which are then joined using binary joins. For the queries we study in this paper, EH, similar to BiGJoin, uses a pure GJ plan based simply on attribute ordering.  EH is highly optimized for evaluating queries on static graphs, and spends a non-trivial amount of time preparing its indices, which vary their representation in response to structural properties of the underlying data.

To guarantee a fair comparison with EH, we run its experiment using the AMI machine provided by the EH team. We started by using a machine with similar configuration as our cluster machines\footnote{An r3.8xlarge AWS machine with 244 GB memory and 32 cores.}, however EH ran out of memory when running the triangle query on TW. Therefore, we used an x1.16xlarge AWS machine with 64 cores and 976 GB memory. We used TW and LJ and the triangle and diamond queries. Unfortunately, EH ran out of memory on the diamond query on TW. 

Table~\ref{table:EmptyHeaded} reports two metrics for both systems: (1) the runtime; and (2) the time to index the input data. 
As shown in the table, our implementations perform worse than EH due to our lack of specific optimizations for static datasets, such as compacting dense extension lists into bit vectors. In exchange, we are able to distribute across multiple machines and respond to changes in input, but this generality comes at a price. We are also evaluating EH's index build time and memory footprint, something EH is explicitly not optimized for, which combined with a lack of distribution limits our ability to evaluate EH on the largest datasets.

\begin{table}
\centering
\begin{tabular}{|p{1.7cm} | p{0.85cm} | p{0.85cm}  | p{1.6cm}  | p{1.5cm}  |}
\hline
Query & EH-R & EH-I & BiGJoinT-R & BiGJoinT-I \\
\hline
\cline{1-5}\cline{1-5}\cline{1-5}
Triangle-LJ & {\bf $1.2$s} & {\bf $150.3$s}   & $6.5$s & $1.9$s \\
\cline{1-5}\cline{1-5}\cline{1-5}
Diamond-LJ & {\bf $31.7$s} & {\bf $150.3$s}   & $712.3$s & $1.9$s \\
\cline{1-5}\cline{1-5}\cline{1-5}
Triangle-TW & {\bf $213.8$s} & {\bf $4155$s}  & $588$s & $34.4$s \\
\cline{1-5}
\end{tabular}
\caption{Comparison against EmptyHeaded. ``-R'' and ``-I'' indicate runtime and index time, respectively. EmptyHeaded's absolute performance is better on a single machine. The index building time can be non-trivial.}
\label{table:EmptyHeaded}
\vspace{-10pt}
\end{table}

\subsubsection{Arabesque}
\label{subsec:arabesque}

Arabesque is a distributed system specialized for finding subgraphs in large graphs. In Arabesque, each distributed worker gets an entire copy of the graph and starts extending a partition of the vertices to form larger and larger subgraphs that are called {\em embeddings}, equivalent to prefixes in our terminology. In Arabesque, prefixes are extended by considering the neighbors of individual vertices, rather than by intersecting the neighborhoods of multiple vertices as GJ does, and correspond to an edge-at-a-time strategy in our terminology . This puts it at a disadvantage for purely structural queries of the sort we examine (though, it more cleanly supports queries like ``subgraphs with average edge density at least 1/2'').

We used Arabesque's most recent version (1.0.1-BETA) which runs on Giraph. 
On our cluster, Arabesque was only able to load the LJ dataset and ran out of memory on our other datasets.
We used the triangle and 4-clique queries. We used 8 machines, each running one Arabesque worker, and each worker using 16 cores. We measured both run-time and intermediate prefixes considered by the system. We used the triangle and 4-clique code provided by the authors of the system but improved the code to not output any intermediate prefixes or final output.\footnote{We note that this code used {\em VertexInducedEmbeddings} of Arabesque, which extend prefixes by one vertex but internally by considering each edge separately.} We repeated the same experiments with BiGJoinT on the same configuration, so using 8 machines with 16 Timely workers on each. 

Table~\ref{table:arabesque} reports the running times as well as the number of intermediate results considered, which partly explain the running times. Arabesque considers roughly 30x more prefixes as BiGJoinT, which manifests as between 10x and 20x higher running times.

\begin{table}
\centering
\begin{tabular}{|p{1.1cm} | p{1.1cm} | p{1.1cm}  | p{1.6cm}  | p{1.5cm}  |}
\hline
Query & Arbsq-R & Arbsq-I. & BiGJoinT-R & BiGJoinT-I \\
\hline
\cline{1-5}\cline{1-5}\cline{1-5}
Triangle & 69.0s & 1.46B & {\bf 3.4s} &  {\bf 38M}  \\
\cline{1-5}\cline{1-5}\cline{1-5}
4-clique & 273.7s & 18.7B & {\bf 21.8s} &  {\bf 350M}  \\
\cline{1-5}
\end{tabular}
\caption{Comparison against Arabesque. ``-R'' and ``-I'' indicate runtime and index time, respectively. BiGJoinT is faster and considers fewer candidate subgraphs than Arabesque.}
\label{table:arabesque}
\vspace{-10pt}
\end{table}

\subsection{Capacity and Scaling}
\label{subsec:capacity-scaling}
When a graph fits in the memory of a single machine, the naive parallelization strategy of replicating the graph to each machine should work very well in practice. That is why one of our primary goals was to scale to graphs (and datasets) whose collected indices do not fit in the memory of a single machine, which our algorithms achieve by using a  working set that is only linear in the input relations.
At the same time, very large graphs can contain prohibitively many instances for even the simplest queries. For example, we estimate that there are over 9 trillion triangles and 23 quadrillion $(2.3 \times 10^{16})$ diamonds in Common Crawl\footnote{These estimates are based on the number of triangles and diamonds we find per edge in our incremental experiments, which are 143 and over 368K, respectively.}. Our goal is therefore not to evaluate BiGJoin when computing all subgraph instances, but Delta-BiGJoin's throughput and capacity when maintaining these queries under updates.

We use the Common Crawl dataset, which has 64B edges and is roughly 1,000x larger than LJ, 50x larger than TW, and 20x larger than UK. When each node ID requires 4 bytes, the graph requires $\approx$512GB written as a list of edges $(u, v)$, and $\approx$ 256GB as an adjacency list. Since we index edges in both directions, our implementation requires $\approx$512GB.

We load up various fractions of the edges in the graph, ranging from one-sixteenth to all edges, and evaluate Delta-BiGJoin on a range of one to sixteen machines. We use 14 workers per machines. So our number of cores/workers range from 14 to 224. Each subset of the graph results in a scaling curve as we increase the number of machines, and we require an increasing number of machines to start the experiments as the size of the subsets grow. For each configuration, we track the number of edges indexed on each machine, the peak memory required, and the throughput of changes (both input and output). We use the triangle query.

\begin{figure}[!t]
			\centering
	\begin{subfigure}{0.47\textwidth}
		\centering
		\includegraphics[height=.45\textwidth,width=1\textwidth]{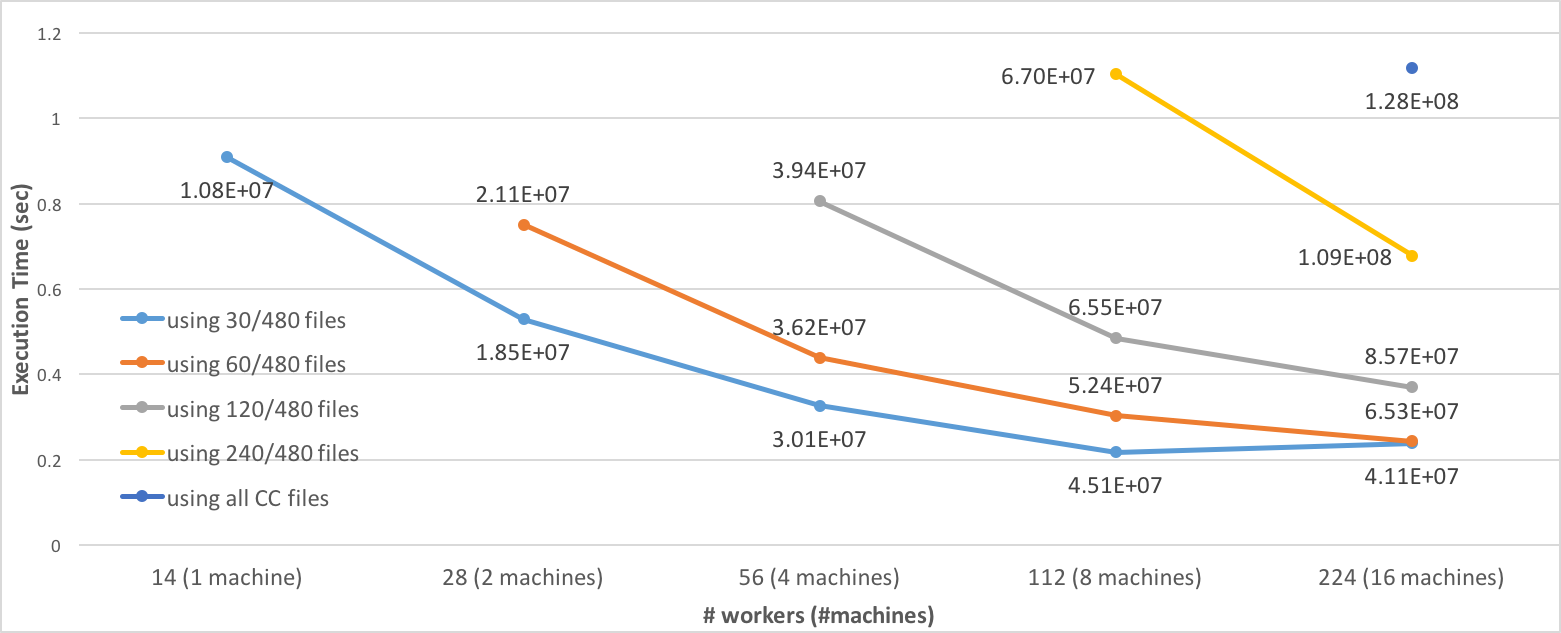}
		\caption{Execution Time; data points are the times to perform a batch of one million updates, averaged across twenty batches. The numbers by each data point report the number of output changes per second (triangles changed). The computation processes roughly 1M updates per-second, reporting between 10M and 100M changed triangles per second.}
		\label{fig:CC-execution-scalability}
	\end{subfigure}
	\begin{subfigure}{0.47\textwidth}
		\centering
		\includegraphics[height=.45\textwidth,width=1\textwidth]{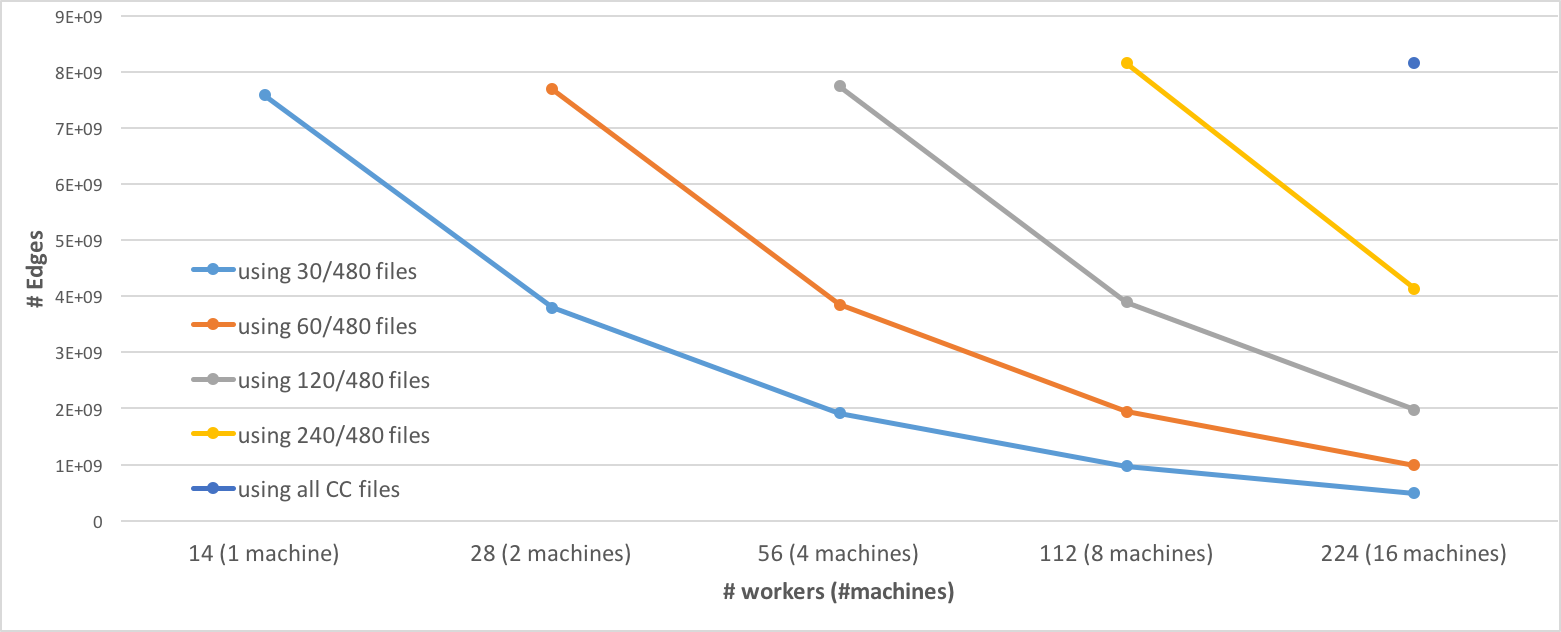}
		\caption{Maximum Index Size per Machine, in total index tuples per machine. Index size decrease roughly linearly with additional machines at each scale.}
		\label{fig:CC-index-size-scalability}
	\end{subfigure}
	\begin{subfigure}[t]{0.47\textwidth}
		\centering
		\includegraphics[height=.45\textwidth,width=1\textwidth]{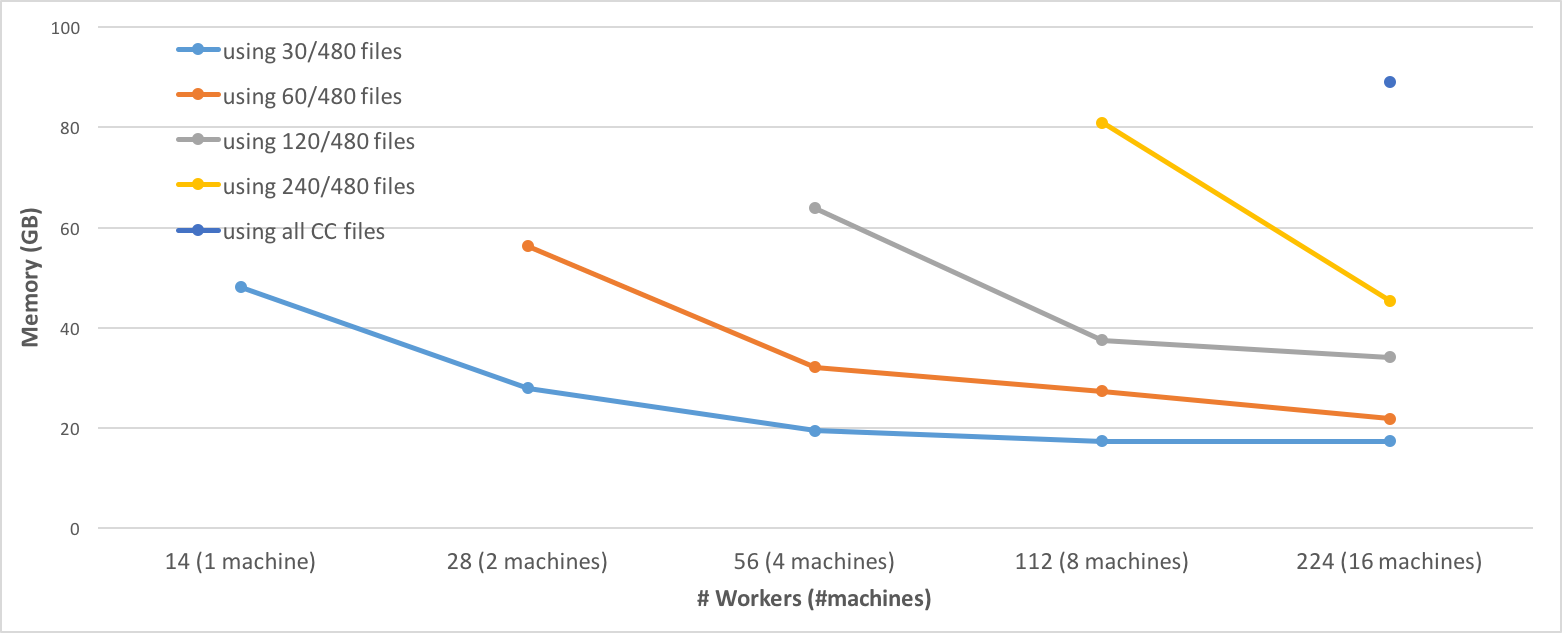}
		\caption{Maximum Memory per Machine, in gigabytes per machine. This peak occurs in initial index building rather than steady-state execution. The maximum does increase as we double the workers and input size, but this appears to be due to execution skew in data loading.}
		\label{fig:CC-memory-scalability}
	\end{subfigure}

	\caption{Scaling as we increase machines (and workers) and the initial graph input. Each line represents an experiment where we pre-load an indicated fraction of the CC dataset, and then perform 20 rounds of 1M input edge updates for a triangle-finding query.}
	\label{fig:CC-scalability}
	\vspace{-15pt}
\end{figure}

Figure~\ref{fig:CC-scalability} shows our scaling results on this large graph. For each fixed subset of the graph, additional workers both improve the thro-ughput and reduce the per-machine index size and memory requirements. The plot of maximum index size (across all machines) indicates that as we double the amount of data and number of workers, the maximum size stays roughly fixed at 8 billion, which is roughly equal to the total tuples divided by the number of machines, indicating effective balance despite some vertices with very high degree (the largest out-degree is  $\approx$45 million). With the exception of the smallest dataset on the largest number of machines, throughput increases and peak memory requirements decrease with further machines; however, as the work gets progressively more thin (one sixteenth of the graph spread across 224 workers) system overheads begin to emerge.

We also report our throughput and the peak memory required when running the diamond and 4-clique queries when loading the full graph and using 224 workers in Table~\ref{table:common-crawl}. Here we see substantially lower throughput of input changes. For example, computing the triangles of a batch of 1M edges with 224 workers after loading the entire graph takes about 1.1 seconds (shown as the highest singleton point on the right in Figure~\ref{fig:CC-execution-scalability}). In contrast, computing the 4-cliques on 200K edges in the same set up takes 226 seconds. However, we see a relatively similar throughput for output changes, in tens of millions, in both cases. That is, each input edge changed results in substantially more subgraph matches changed, and it is the volume of output that limits our throughput.

\begin{table}
\centering
\begin{tabular}{|p{1.5cm} | p{1.6cm} | p{2cm} | p{1.5cm} |}
\hline
Query & Average Time / batch& Output Throughput & Max. Mem.\\
\hline
\cline{1-4}\cline{1-4}\cline{1-4}
4-clique & {\bf $226.378$ s} & {\bf $46,517,875$ /s}  & {\bf $108.4$ GB} \\
\cline{1-4}
Diamond & {\bf $276.587$ s} & {\bf $26,681,430$ /s}  & {\bf $92.6$ GB} \\

\cline{1-4}\cline{1-4}\cline{1-4}

\end{tabular}
\caption{Common Crawl experiments. Sixteen machines load 64 billion edges, index them, and track motifs in $20$ batches of $10K$ random edge changes. Although the input throughput is much lower than for triangles, the {\em output} throughput remains relatively high at tens of millions of observed subgraph changes per second.}
\label{table:common-crawl}
\vspace{-15pt}
\end{table}

\subsection{Generality and Specializations}
\label{subsec:seed}

In this section we show that our algorithms can employ existing optimizations from subgraph queries and multiway joins literature. In doing so, we also achieve two things. First, we compare our work to the recent SEED~\cite{lai:seed} work, which develops efficient optimizations for evaluating undirected subgraph queries in the distributed setting. Second, by implementing one of the optimizations, we demonstrate that our approach can take as input general relations instead of the binary {\code edge($a_i$, $a_j$)} relations we used so far. We implement the following three optimizations:
\squishlist
\item {\bf Symmetry Breaking:} SEED imposes constraints on vertex IDs to break symmetries. For example, for \texttt{4-clique} query, we might constrain that $a_1 < a_2 < a_3 < a_4$.
This allows finding each undirected four-clique once instead of 24 times, for each permutation of the vertices in the clique. One can be more efficient by first ordering by \emph{degree}, and then by ID if there are ties. This is commonly accomplished by giving new IDs to the vertices so that they are ordered by degree, and edges point from vertices with lower ID to higher ID. We incorporate this optimization by transforming the input dataset, and supporting inequality constraints (which are just filters applied to intermediate prefixes).
\item {\bf Triangle Indexing:} SEED builds index structures over small non-trivial subgraphs, such as triangles. These indices provide more direct access to relevant vertex IDs reflecting multiple constraints already imposed. The ideas are similar to the recent FAQ work~\cite{abokhamis:faq}, which identifies some common subqueries in larger queries (for example, triangles in a four-clique query) and materializes these subqueries. We incorporate this optimization by first finding all the triangles in the graph and then writing these as a ternary relation \texttt{tri}$(a_i, a_j, a_k)$. Since we support general relational queries and can index general relations, we can index \texttt{tri}$(a_i, a_j, a_k)$ by $(a_i, a_j)$ and provide efficient random access to vertices $a_k$ that complete a triangle with $(a_i, a_j)$. 
Using the \texttt{tri} relation, \texttt{4-clique} query simplifies to:
$$ \texttt{tri}(a_1,a_2,a_3), \texttt{tri}(a_1,a_2,a_4), \texttt{tri}(a_1,a_3,a_4). $$ 
This rewriting reduces the complexity of the query, and results in fewer intermediate prefixes explored. We stress that this is not precisely the same optimization SEED does. SEED indexes triangles by $a_1$ so that full neighborhoods of each vertex is available, revealing large cliques at once. Our optimization is closer in spirit to the FAQ work, but demonstrates the utility of supporting general relations in evaluating subgraph queries.
\item {\bf Factorization:} The \texttt{house} query is amenable to a technique called {\em factorization}~\cite{olteanu:factorized-dbs}, which expresses parts of the query results as Cartesian products. In the \texttt{house} query, $(a_2, a_3, a_4, a_5)$ form a clique and the missing edges are $(a_1, a_4)$ and $(a_1, a_5)$. We can first compute the triangle $(a_2, a_3, a_4)$ and then perform two independent extensions to the lists of $a_1$ and $a_5$ values. As these two variables do not constrain each other, they can be left as lists rather than flattened into the list of their Cartesian product. 
The SEED work proposes a similar optimization (named SEED+O) in which large cliques are kept as cliques, rather than explicitly enumerating all bindings to variables. We use this optimization only for the \texttt{house} query.

\end{list}

Table~\ref{table:SEED} compares SEED+O (SEED with clique optimizations) measurements taken from their paper with three variations of our work: (i) vanilla BiGJoinT, (ii) BiGJoinT with symmetry breaking (BiGJoinT-SYM), and (iii) BiGJoinT with symmetry breaking and triangle indexing (BiGJoinT-SYM-TR). All of our \texttt{house} measurements also contain the factorization optimization. We used 10 machines with 16 cores, which is a cluster setup similar to the one used in the SEED paper. Table~\ref{table:SEED} demonstrates two things: (1) Our algorithms have the flexibility to employ several optimizations from prior work to become more efficient; and (2) The results of the 4-clique and 5-clique queries demonstrate that we are initially competitive with SEED using the same resources, and when incorporating some of their optimizations, we can even outperform it. We emphasize that the SEED measurements are reported from~\cite{lai:seed} rather than reproduced on identical hardware, and that our goal is not to provide evidence that our work outperforms SEED so much as that our work is able to accommodate similar optimizations.

\begin{table}
\centering
\begin{tabular}{|p{1.3cm} | p{1.3cm} | p{1.3cm}  | p{1.3cm}  | p{1.3cm}  |}
\hline
Query & SEED-O & BiGJoinT & BiGJoinT-SYM & BiGJoinT-SYM-TR \\
\hline
\cline{1-5}\cline{1-5}\cline{1-5}
4-clique & 60s & 54.0s & 43.4s & 13.3s \\
\cline{1-5}\cline{1-5}\cline{1-5}
house & 1013s & 370.0s & 294.3s & 74.1s \\
\cline{1-5}\cline{1-5}\cline{1-5}
5-clique & 1206s & 2861.1s & 2153.2s & 315.7s \\
\cline{1-5}
\end{tabular}
\caption{Comparison with SEED, against three BiGJoin variants including several optimizations: breaking symmetry by renaming vertices by degree (-SYM) and then re-using pre-computed triangles (-TR). BiGJoin's absolute performance is comparable to optimized approaches, and improves as optimizations are applied.}
\label{table:SEED}
\end{table}

\subsection{Sensitivity to Batch Size}
\label{subsec:batch}
We finally evaluate the effects of the batch size on our algorithms. Batch size affects two aspects of our algorithms. First, very small batch sizes can impede parallelism. As an extreme example, consider finding all instances of a subgraph in a graph with a batch size of 1. Then at least initially only one worker in the cluster will do count minimization, candidate proposals, and intersections.  Second, with larger batch sizes, we expect the algorithm to use more cluster memory. Therefore we expect that as batch sizes get larger, runtime improves because the algorithm can parallelize better but after we get to a large enough batch size, we expect the algorithm to have a stable runtime but use more memory. 

\begin{figure}[!t]
\centering
\includegraphics[width=0.4\textwidth]{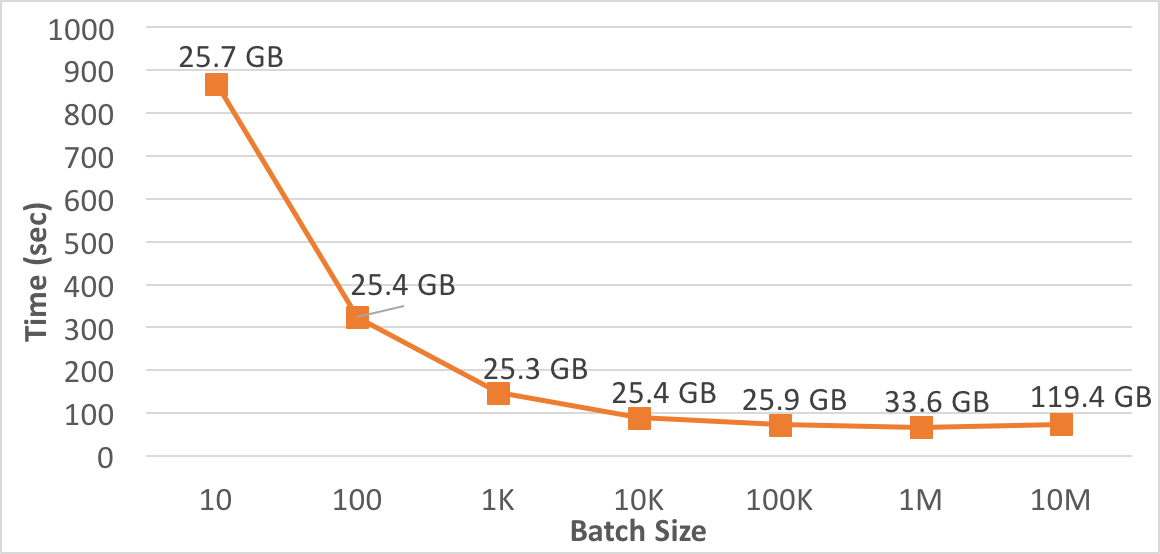}
\caption{Effects of batch size. Note that the maximum memory usage in small batches is very close to the index size ($25.1$ GB).}
\label{fig:batch-size}
\end{figure}

To test this, we used the triangle query and ran Delta-BiGJoin on the UK graph using 16 workers on 1 machine and using batch sizes of 10, 100, 1K, 10K, 100K, 1M, and 10M. We first load the dataset, then ran Delta-BiGJoinT using 10M edges. The results are shown in Figure~\ref{fig:batch-size}. The numbers on top of the points indicate the maximum memory usage\footnote{We measure memory usage using an operating system tool which reports a snapshot of memory usage every second instead of the average memory usage every second. This explains the small approximation and inaccuracy in the reported memory size.}. As shown in the figure, indeed as batch size increases the runtime initially improves and then remains the same around after batch size of 10K. As we expect, larger batch sizes lead to more cluster memory usage. We note that the increase in the memory usage is very small for batch sizes less than or equal to 100K because the intermediate data that the algorithm generates with these batch sizes is insignificant compared to the size of the input graph. Batch size is a useful parameter to balance memory usage and speedup.

\section{Related Work}
\label{sec:rw}
We reviewed related work on worst-case join algorithms in Section~\ref{subsec:wcja}. Here, we review related work in distributed join algorithms, incremental view maintenance, graph processing systems, algorithms that evaluate one-time subgraph queries, and streaming or semi-streaming algorithms for subgraph finding.

{\bf {\em Distributed Multiway Join Algorithms}}: We reviewed some of the algorithms based on the Shares algorithm in Section~\ref{subsec:shares}. A more recent algorithm~\cite{hu:joins} has introduced a new distributed algorithm for queries that involve only two relations based on sorting the relations on their join attributes. This contrasts with the hashing approach of Shares. The algorithm runs for a small number of rounds, requires cumulative memory and communication that is as large as the actual output but does not generalize to more complex joins, e.g., involving three relations.

Reference~\cite{afrati:gym} has introduced a multiround join algorithm called {\em GYM}, which takes as input a {\em generalized hypertree decomposition} (GHD) $D$ of $Q$. The algorithm first computes several intermediate relations based on $D$ in one round using Shares. Then the algorithm runs a distributed version of Yannakakis's algorithm for acyclic queries~\cite{yannakakis:acyclic}. Overall the algorithm runs for $O(n)$ rounds and incurs a communication and cumulative memory cost of $O(\IN^w + \OUT)$, where $w \ge 1$ is called the {\em width} of the GHD $D$ and $\OUT$ is the actual output size. This amount of communication cost is always $O(MaxOut_Q)$ but $w$ is only $1$ for acyclic queries, so for any cyclic query the memory requirements of GYM is superlinear in $\IN$.

Reference~\cite{joglekar:degree} introduces another algorithm, which we refer to as {\em the DBP algorithm}. DBP algorithm takes 3 rounds and takes $O(L\times \IN^{DBP(L)} + \OUT)$, where $L$ is a free parameter that indicates load per machine and $DBP(L)$ is called the {\em degree-based packing bound} of the query for load $L$. Similar to GYM, for any $L$, DBP's communication is always $O(MaxOut_Q)$ but (for any $L$) the algorithm can require a cluster memory that is superlinear in $\IN$ as it computes  intermediate relations that can be superlinear in $\IN$.

{\bf {\em Incremental View Maintenance (IVM)}}: There is a vast body of work on incrementally maintaining views that contain selection, projection, joins, group-by-aggregates, among others. We refer the readers to reference~\cite{chirkova:ivm} for a survey of these techniques. The overall technique of Delta-BiGJoin falls under the {\em algebraic technique} of representing updates to tables as delta relations and maintaining views through a set of relational algebraic queries. This approach has been extensively studied by previous work. Prior work on algebraic techniques range from addressing limitations of delta query-based techniques, e.g., when evaluating a top-k query~\cite{yi:top-k}, 
to techniques using higher-delta queries~\cite{ahmad:dbtoaster}, e.g., delta queries of delta queries of a query. When evaluating subgraph queries, these techniques do not yield theoretically optimal results and may require materializing very large intermediate results. 

The only IVM algorithm with known theoretical guarantees and the one closest to our work is the algorithm described in reference~\cite{veldhuizen:inc-lftj}. This IVM algorithm is based on the Leapfrog TrieJoin (LFTJ) worst-case optimal join algorithm. We refer to this algorithm as LFTJ-Inc. Similar to GJ, LFTJ  is based on doing intersections of multiple extension sets in time proportional to the size of the minimum-size set. Unlike our description of GJ, which uses hash-based indices, LFTJ uses tries to index the prefixes of the tuples in each input relation. 
LFTJ-Inc uses another set of indices called {\em sensitivity indices} which, for each prefix $p$, store the set of intervals in the extensions of $p$ such that any update to these intervals {\em could result} in the output of the query to change. For example, consider a join $R(a_1, a_2) \bowtie S(a_2, a_3)$.  A sensitivity index for $R$ could store $(5, [-\infty, 8))$, meaning that if a tuple with $a_1=5$ and $a_2 \in [-\infty, 8)$ is added or deleted from $R$, this update could change the output of the join.  Using the sensitivity indices, LFTJ-Inc ``fixes'' the necessary intersections to compute the outputs that have changed. Between any two updates, LFTJ-Inc maintains query results in time proportional to what the author calls the {\em trace edit distance} of running LFTJ on the relations before and after the update. That is, the author analyzes the ``trace'' of LFTJ, which is the set of iterator operations that the algorithm does, on inputs before and after the update, and conclude that the work that LFTJ-Inc does to maintain the query result is proportional to the amount of work one would need to ``fix'' LFTJ's iterator operators before the update. We note that this is a stronger theoretical guarantee than our Delta-BiGJoin's worst-case optimality under insertion-only workloads. In particular a trace edit distance guarantee implies that LFTJ-Inc is worst-case optimal under insertion-only workloads. However unlike Delta-BiGJoin, which requires indices linear in the input sizes, the sensitivity indices could be as large as the AGM bound of the query (so super-linear in the size of the input) for some queries and thus require a prohibitively large amount of memory.

{\bf {\em Systems and Algorithms For One-time Subgraph Queries}}:
 Although they significantly differ in their graph storage, algorithms, and optimizations, existing systems that evaluate general subgraph queries are based on the edge-at-a-time strategy, unlike BiGJoin's vertex-at-a-time strategy. We reviewed Arabesque, EmptyHeaded, and SEED in Sections~\ref{subsec:arabesque},~\ref{subsec:eh}, and~\ref{subsec:seed}. We review other work below.
  
{\bf PSgL~\cite{shao:psgl}:} PSgL is a subgraph enumeration system that is built on top of Giraph~\cite{giraph}. PSgL picks an order of the vertices (i.e., attributes), say $a_1, ..., a_m$ in $Q$, called a {\em traversal order}. It starts with candidate partial matches $G_{psi}$ for $a_1$, then extends each $G_{psi}$ to all neighbors of $a_1$ in $Q$ (not just $a_2$). When matching $a_j$, the existence of edges $(a_i, a_j)$ edges for $i$ $<$ $j$ will be checked and if they exist $a_j$ will be extended to all neighbors $a_k$ $>$ $a_j$. This is effectively an edge-at-a-time strategy. The paper presents techniques for picking good traversal orders, balancing workload among workers, and breaking internal symmetries in queries over undirected graphs, which can complement our algorithms on undirected graphs as well.

TrinityRDF~\cite{zeng:trinity-rdf} and Spartex~\cite{abdelaziz:spartex} are two distributed RDF engines that can evaluate any SPARQL~\cite{sparql} query. SPARQL queries can express any subgraph query, so both of these systems can evaluate general subgraph queries. The optimizers of both systems use edge-at-a-time strategies although they use different techniques to choose edge extension plans.

There are several other work, such as references~\cite{arifuzzaman:patric, park:triangles, wu:subgraphs} that  describe data distribution techniques or other optimizations to find subgraphs in a distributed setting, using black box or naive subgraph finding algorithms as subroutines. We do not review these references here. 
  There are also several studies that study evaluation of a single specific query, e.g., the triangle query~\cite{dolev:triangles, gupta:diamond}, which we do not review here.

{\bf {\em Streaming and Semi-streaming Algorithms}}: 
Several works stu-dy variants of continuous subgraphs queries in a streaming or semi-streaming setting, i.e., in which the algorithms can use slightly superlinear space. A thorough review of these works is beyond the scope of our work. Example studies include those that focus on triangle finding and variants~\cite{becchetti:triangle, tangwongsan:triangle}.
Many works in this area focus on approximating the counts of different subgraphs and instead of enumerating, which is the problem we study in this paper.

\section{Future Work}
\label{sec:fw}

We outline three broad directions for future work. First is studying the extent of workload imbalance in real-world graphs and designing more efficient workload-balanced versions of BiGJoin. Although BiGJoin-S is theoretically skew-resilient, in our preliminary implementation of the algorithm, we observed that its overheads were higher than its benefits. Better understanding the effects of skew, when it hurts BiGJoin and DeltaBiGJoin's performance, and how to effectively guard against it is an interesting future direction. Second, we are interested in studying how to utilize internal symmetries of queries during query evaluation. For example, when evaluating the 4-clique query, some of the delta-queries, e.g., $dQ2$ and $dQ3$, compute the same $P_2$ and $P_3$ prefixes due to internal symmetry of the query. An interesting future direction is to automatically exploit such symmetries to share computations across multiple dataflows of delta queries.
Finally, from a theoretical perspective, an interesting direction is designing practical algorithms that have stronger guarantees than worst-case optimality. A stronger than worst-case optimality guarantee could be optimality in terms of {\em certificate complexity}, which is achieved by the recent serial {\em Minesweeper} algorithm for multiway joins in terms of computation cost~\cite{ngo:minesweeper}. At a high-level, certificate complexity captures the smallest proof size to verify that the output is correct and is a strictly stronger notion than worst-case optimality. 

\section{Acknowledgements}

We would like to thank Lori Paniak for assisting with numerous systems issues. We would also like to thank Michael Isard and Chris R{\'{e}} for early discussions that resulted in the first implementation of BiGJoin. 

\balance
\small{
\bibliography{distributed-motif-detection}
}
\iftechreport
\appendix
\section{Example Execution of GJ}
\label{app:gj-example}
\begin{example} 
\label{ex:running-ex}
Consider evaluating the triangle query \linebreak $Q(a_1, a_2, a_3) := R_1(a_1, a_2)$, $R_2(a_2, a_3), R_3(a_3, a_1)$, where each $R_i$ is an exact replica of the edges in the input graph in  Figure~\ref{fig:ex-input-graph}. GJ executes as follows:
\squishlist
\item  $P_0$ and $P_1$: GJ starts with $P_0=\{ \epsilon \}$, and extends $\{ \epsilon \}$ to $P_1 = \epsilon \times (Ext^1_1[\epsilon] = \{ 1, 2, 3, 4, 5, 6, 7 \} \cap Ext^3_1[\epsilon] = \{ 1, 6, 7, 8, 9,$ $10, 11 \})$. $Ext^1_1$ and $Ext^3_1$ correspond to indices over $R_1$ and $R_3$, respectively.
In this case, neither set is smaller than the other and GJ is free to choose arbitrarily. This intersection produces $P_1 = \{ (1), (6), (7) \}$. 

\item $P_2$: GJ extends each prefix $p$ in $P$ with valid $a_2$ producing $p \times (Ext^1_2[p] \cap Ext^2_2[p])$. This is done by considering the sizes of $Ext^1_2[p]$ and $Ext^2_2[p]$, which for $(1)$ are $Ext^1_2[(1)] = \{ 6 \}$ and $Ext^2_2[(1)] = \{ 1, 2, 3, 4, 5, 6, 7 \}$.
The former index is smaller, and so GJ starts from the set $\{ 6 \}$ and intersects it with $Ext^2_2[(1)]$, producing $(1, 6)$. Other extensions in $P_2$ are $(6, 7)$ and $(7, 1)$.

\item $P_3$: Finally, GJ extends each of these three prefixes using $Ext^2_3$ and $Ext^3_3$, again starting from the smaller of the candidate extensions for each prefix. For example, when extending $(1, 6)$, $Ext^2_3 = \{ 7, 8, 9, 10, 11 \}$ is intersected with $Ext^3_3 =  \{ 7 \}$ giving the triangle  $(1, 6, 7)$. Similarly $(6, 7)$ and  $(7, 1)$ give the outputs $(6, 7, 1)$ and $(7, 1, 6)$, respectively. 
\end{list}
\end{example}
\begin{figure}[t!]
\centering
  \includegraphics[width=.4\linewidth]{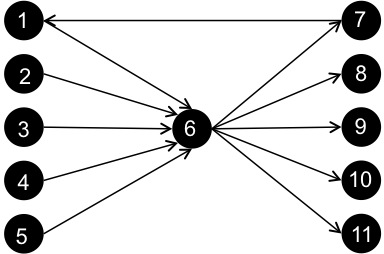}
\vspace{-5pt}
  \caption{Example input graph.}
\label{fig:ex-input-graph}
\vspace{-15pt}
\end{figure}

\section{Proof of Lemma 3.1}
\label{app:thm-bigjoin}

Recall that each {\code Proposal} operator stores $(p, c, i, rem\text{-}ext)$ quadruples, where {\em rem-ext} is the metadata the operator keeps for prefix $p$ to track the remaining amount of candidate extensions the operator has to do for $p$. We call this metadata the remaining {\em intersection work for $p$}. We assume each {\code Proposal} operator, for each $P_j$ keeps track of the {\em cumulative intersection work} it has to do for the set of prefixes it has in $P_j$. Figure~\ref{fig:bigjoin-stages} shows the stages of BiGJoin's dataflow, where unlike Figure~\ref{fig:dataflow-primitive} we start each dataflow from the {\code Proposal} operator and omit the dataflow extending $P_0$ to $P_1$. As discussed in Section~\ref{sec:bigjoin}, BiGJoin picks a $P_j$ to extend to $P_{j+1}$ prefixes, where each operator extends a subset of its prefixes up to at most $B'$ extensions. Recall that BiGJoin picks the $P_j$ with the largest $j$ value (where j is from 1 to $m-1$) such that at least one {\code Proposal} operator has $B'$ cumulative intersection work to do. We let $P_{j^*}$ be the prefix set BiGJoin picks.  If $j^* < m-1$ then the algorithm generates a batch of $P_{j^*+1}$ prefixes. Otherwise if $j^* = m-1$, it produces and writes a batch of outputs. We assume throughout our analysis that the operators of the dataflow run across different workers and we use the terms operator and worker interchangeably. Moreover there is a synchronization between any two operators (not only the {\code Proposal} but also {\code Count} and {\code Intersect}) in the dataflow consisting of a round in MPC terms. This is actually needed in order to analyze our dataflow in MPC because in a single round of MPC, workers cannot perform computations on the data they receive. Therefore we assume there is a synchronization barrier between each arrow in Figure~\ref{fig:bigjoin-stages}.

We observe that BiGJoin maintains two invariants: (1) At any point in time, each {\em Proposal operator} has at most $2B'$ prefixes (not candidate extensions) from each $P_j$; and (2) at each round of computation the amount of intermediate data due to candidate extensions being intersected is at most $O(wB')$. Initially both invariants hold because only one {\em Proposal operator} has the $P_0=\{(\epsilon)\}$ prefix and its quadruple and other workers do not hold any prefixes. 

At any point in time, since each {\code Proposal} operator extends at most $B'$ candidates, and that the {\code Count} and {\code Intersect} operators do not generate more data than their inputs, and that there are $w$ workers, the amount of intermediate data is bounded by $wB'$, so the second invariant is satisfied. Note that the first invariant is also maintained because each {\code Proposal} operator $w_i$ generates at most $B'$ new $P_{j^*+1}$ prefixes for itself. However, note that $w_i$ must have less than $B'$ $P_{j^*+1}$ prefixes, because otherwise BiGJoin would have decided to extend $P_{j^*+1}$ prefixes instead of extending $P_{j^*}$ prefixes. This proves that the cumulative memory BiGJoin needs to store all of the $P_j$ prefixes is $O(mwB')$.

We next analyze BiGJoin's communication and computation costs. Note that cumulatively BiGJoin performs exactly the same amount of computation as GJ. To see this, first note that we assume that the constant counting and test membership assumptions hold when {\code Intersect}, {\code Proposal} and {\code Count} operators perform appropriate operations on the $Ext_j^i$ indices. Second, similar to GJ, for each prefix $p$, BiGJoin starts its intersections from the relation that has the minimum number of extensions. Even though BiGJoin can intersect the extensions of each prefix $p$ in multiple batches, effectively each of the possible extensions gets intersected with at most $n$ different relations incurring a computation cost of at most $n$. GJ similarly perform exactly same number of intersections for each of the candidate extensions of $p$. In addition, BiGJoin incurs an equivalent amount of communication cost when doing the intersections of each of the candidate extensions. This is because for each intersection the candidate extension is sent to another operator.  
Therefore, the cumulative computation and communication cost of BiGJoin is the same as the computation cost of GJ, which is $O(mnMaxOut_Q)$. For memory consumption, we showed that the two invariants above are satisfied. Therefore, beside the indexing cost, each worker holds at most $2B'$ $P_j$ prefixes for each $j$. Therefore, the cumulative memory needed to store all of the $P_j$'s is $O(mwB') = O(mB)$. 

Finally, we analyze the number of rounds of computation BiGJoin takes in MPC terms. There are at most $2n+1$ operators in each dataflow primitive extending $P_j$ to $P_{j+1}$, as there are at most $n$ {\code Intersect} and $n$ {\code Count} operators. Therefore an iteration of extending $B'$ prefixes from one of the $m$ $P_j$ sets takes at most $2n+1$ rounds. Note that by the AGM bound, the size of each $P_j$ set is at most $MaxOut_Q$. Therefore the number of rounds of computation is $O(\frac{mnMaxOut_Q}{B'})$, completing the proof.

\section{Proof of Theorem 3.2}
\label{app:thm-deltagj}
Throughout this section we consider a query $Q$ and a series of updates, $\Delta(1), \Delta(2), ...$, that modify the input relations $R_i$ of $Q$. We denote by $\Delta R_i(z)$ the set of updates to $R_i$ in $\Delta(z)$. We denote by $R_i(z)$ the state of relation $R_i$ after incorporating all of the updates until and including $\Delta R_i (z)$, i.e., $R_i(z) = \Delta R_i(1) + \Delta R_i(2) + ... + \Delta R_i(z)$. 

\begin{figure}[t!]
  \centering
	\includegraphics[width=0.47\textwidth]{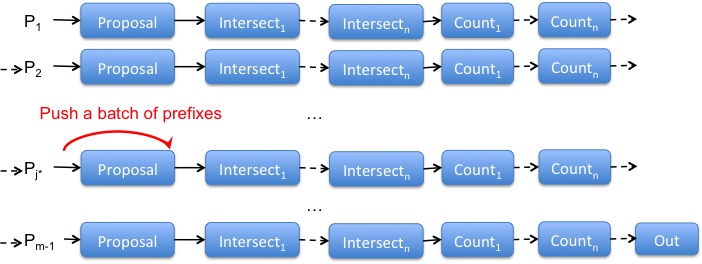}
  \vspace{-10pt}
  \caption{BiGJoin Dataflow.}
  \label{fig:bigjoin-stages}
  \vspace{-15pt}
\end{figure}

Delta-GJ evaluates each $dQ_i$ $z$ times, one for each update. For each $dQ_i$ we show that the cumulative cost of these $z$ evaluations is asymptotically at most the running time of $GJ$ on $R_i(z)$ with the same attribute ordering Delta-GJ uses for $dQ_i$. We let the query that GJ evaluates on the final relations as $Q(z) := R_1(z), ..., R_n(z)$. Note that by Theorem~\ref{thm:gj}, the cost of running GJ with any attribute ordering is $O(mnMaxOut_Q)$. The claim of the theorem then follows from the fact that there are $n$ delta queries. We need to account for Delta-GJ's three costs when evaluating $dQ_i$ $z$ times: (1) cost of indexing; (2) cost of computing $P_{r_i}$ tuples (recall from Section~\ref{sec:preliminaries} $r_i = |R_i|$); and (3) cost of computing the rest of $P_{r_i + 1}, ..., P_{m}$. 

For cost (1), note that the cost of indexing each tuple $t$ in \linebreak $\Delta R_i(1), ..., \Delta R_i(z)$ first into $r_i$ $Ext^i_j$ indices is exactly the same as the cost of $GJ$'s indexing cost of $t$ when evaluating $Q$ on $R_i(z)$.

For cost (3), note that once $P_{r_i}$ has been computed, the cost of Delta-GJ for extending any tuple $t \in P_{r_i}$ to $P_j$ for $r_i < j \le m$ is clearly less than the cost that GJ incurs when extending $t$ when GJ fixes the same global attribute ordering as Delta-GJ. This is because by definition, across all $z$ insertions, $dQ_i := R_1(z), R_2(z), ..., R_i(z), R_{i+1}(z-1), ..., R_n(z-1)$ and each relation in this query is a subset of its corresponding relation in $Q(z)$

Finally for cost (2), recall that Delta-GJ incurs a cost of $O(mn)$ for each tuple $t \in \Delta R_i$  to verify whether $t$ is in $P_{r_i}$ or not. Cumulatively across all of the $z$ updates, these checks incur an extra cost of $O(mn|R_i(z)|)$. Moreover, cumulatively across all of the relations, this cost is  $O(mn\IN)$. We note that in the best case $MaxOut_Q$ is $\Omega(\frac{\IN}{n})$, since the largest relation is of size at least $\frac{\IN}{n}$ and the worst-case output of any query is at least the size of its largest relation. Therefore $mn^2MaxOut_Q$ is at least $mn\IN$, which implies that  $O(mn\IN)$ is a cost that is subsumed by $O(mn^2MaxOut_Q)$, completing the proof.

\section{Proof of Lemma 3.3}
\label{app:thm-deltabigjoin}
Note that by Thereom~\ref{obs:bigjoin} BiGJoin's communication and computation cost on any query $Q$ is asymptotically the same as the computation cost of running GJ on $Q$. By the same arguments in Theorem~\ref{thm:delta-gj}, the communication and computation cost of Delta-BiGJoin is therefore  $O(mn^2MaxOut_Q)$. Note also that the cumulative memory cost of Delta-BiGJoin is never larger than the  $O(mn\IN(z) + mB)$, which is the memory cost of running BiGJoin on $Q(z)$. This is because at any point in time, the indices that Delta-BiGJoin uses is at most as large as the indices that BiGJoin uses on $Q(z)$. Similarly, the intermediate data that Delta-BiGJoin generates on any $dQ_i$ query is at most as large as the intermediate data that BiGJoin generates on $Q(z)$.  For the number of rounds, note that after each update, Delta-BiGJoin runs $O(mn)$ rounds of computation to compute $P_{r_i}$. From then on, the algorithm will extend prefixes and perform intersections using the same batch size of $B'$ as BiGJoin does. Except, we note that for each update Delta-BiGJoin might run at most another extra $O(mn)$ rounds of computation if the updates are small or if they do not generate enough $P_{r_i +1}, ..., P_m$ prefixes. Note that cumulatively, Delta-GJ, for each $dQ_i$, computes the same $P_{r_i}$ as BiGJoin does on $Q(z)$. Therefore, for each $dQ_i$, cumulatively across all of the $z$ updates, Delta-BiGJoin runs the same number of rounds as GJ does and an additional $O(zmn)$ rounds. Since there are $n$ delta queries, the total number of rounds of computation is  $O(mn^2\frac{MaxOut_Q}{B'} + zmn^2)$, completing the proof.

\section{Proof of Theorem 3.4}
\label{app:thm-bigjoin-s}
First observe that similar to BiGJoin, for each prefix $p\in P_j$, BiGJoin-S does $O(n)$ work in the {\code Count} steps and  $O(n \times min{\text-}c)$ computation and communication in the {\code Extension-Resolve} and {\code Intersect} stages, where $min{\text-}c$ is the minimum count for $p$ after count minimization. That is because there is $O(n)$ work done for each of the $min{\text-}c$ $(p \bullet e)$ candidate extensions of $p$. Note also that the work done in {\code Balance} for $p$ is dominated by the work done in {\code Intersect} steps because for each worker that  $(p, min{\text -}i, {\code start}, {\code end})$ tuples is sent to, there is at least one intersection that will be performed. This proves that the cumulative computation and communication costs of BiGJoin-S and BiGJoin are asymptotically the same.

Next, we prove an invariant of BiGJoin-S that at any point in time the total number of $(p, min{\text-}i,$ $start, end)$ tuples in $P_j$ never exceeds $4B$.
\begin{proof}[of the invariant]
Note that initially the invariant holds. Then suppose that at some point in time BiGJoin-S picks a $j^*$ to start extending $B$ tuples from. This implies that not every worker had at least $B'$ intersection work and therefore $B'$ many $P_{j^*+1}$ tuples (possibly much less since each tuple constitutes possibly multiple intersection work). BigJoin-S then picks a total of $B=wB'$ $(p, k)$ candidate extensions to intersect and then perform count minimization. The {\code Intersect} and {\code Count} operators can only decrease the the number of $B$ candidate extension tuples. At this point there are at most $B$ $(p', min{\text-}i, min{\text-}c)$ triples. Note that the greedy procedure of the {\code Balance} procedure that takes these triples may split a triple $t$ into multiple $(p', min{\text-}i, start, end)$ quadruples, yet each worker $w_{\ell}$ generates at most 1 new triple for each worker $w_{\ell'}$ (only for the very last triple that $w_{\ell}$ gives to $w_{\ell'}$). Therefore cumulatively, this process will generate at most $w^2$ new tuples, $w$ for each worker. Since we assumed $B'\ge w$, so $B \ge w^2$, and extending $B$ $P_j$ tuples generates at most $B + w^2 \le 2B$ $P_{j+1}$ tuples, completing the proof of the invariant.

Therefore the total cumulative memory BiGJoin-S uses is $O(mn\IN + mB)$, where $O(mn\IN)$ is the memory cost of storing the indices and $mB$ is the cost of storing the $P_j$ tuples. 

Next, note that similar to BiGJoin, each batch of $P_j$-to-$P_{j+1}$ computation takes $O(mn)$ rounds. Since BiGJoin-S starts intersecting exactly $B$ candidate extensions in each batch\footnote{In the very last batch of each $P_j$ extension, only once for each $P_j$ there maybe less than $B$ candidate extensions to finish the last}, there are at most $O(\frac{MaxOut_Q}{B})$ batches of $P_j$ that BiGJoin-S does computation on, proving that BiGJoin-S takes $O(\frac{mnMaxOut_Q}{B})$ rounds in total.

Finally, we analyze the load, i.e., per-worker communication, and per-worker computation costs in any batch of BiGJoin-S. We analyze each stage separately:

\noindent {\bf Extension-Resolve}: In the {\code Extension-Resolve} operator, each worker does exactly $B'$ distributed lookups in {\code Ext-Resolve} indices. As we noted earlier, there may be skew in these lookups and in the worst case even all of them could be directed to the same worker. However, the workers guard against this by first aggregating all of their lookups directed to the same key. So for each key each worker gets at most $w$ separate lookup requests. This is an instance of weighted balls-into-bins problem. We show that with high probability each worker sends or receives at most $O(B')$ tuples as follows. After the local aggregation of workers, suppose there are $K$ unique keys left, where let the weight of each key $k_i$ be $z_i$, where $z_i$ is between 1 and $w$. There are two cases: (1) If $K < B'/w$, then we assume for simplicity that all of the keys go to a single worker, so the load of that worker is at most $Kw=B'$. (2) If $K \ge B'/w$, then let $T = \Sigma_i z_i$. We know $B' \le T \le B$ as there are at least $B'$ keys. In this case, consider an adversary that picks the weights of the keys in order to maximize the maximum weight of a bin. If this adversary had the extra flexibility to assign weights of 0 to some of the keys, then reference~\cite{sanders:balls-in-bins} has shown that the worst distribution the adversary can design assigns a weight of $w$ to $T/w$ of the keys and a weight 0 to the rest of the keys (see Lemma 1 in the reference). Even with this extra flexibility, since $T/w \ge B'/w \ge \log(\IN \times MaxOut_Q)$, by Chernoff bounds with probability at least $1-\frac{1}{\IN \times MaxOut_Q}$, each bin gets at most $O(T/w) = O(B/w)$ total weight. Therefore, considering both cases (1) and (2), each worker with probability at least $1-\frac{1}{\IN \times MaxOut_Q}$, receives at most $O(B')$ requests to its {\code Extension-Resolve} indices.

\noindent {\bf Intersect}: By the exact same argument we did for analyzing the load of lookups in the {\code Extension-Resolve} indices, the number of requests each worker sends and receives in each round of {\code Intersect} to $Ext_j^i$ indices (after local aggregation of similar lookups as done in {\code Extension-} {\code Resolve}) is $O(B')$ with high probability. We note that during the intersection, the number of extensions can decrease from $B$, however this can only reduce the load of workers. 

\noindent {\bf Count}: Similarly, in the {\code Count} stage, there are at most $B$ lookups and by the same argument as in {\code Intersect} and {\code Extension-Resolve}, each worker sends or receives at most $O(B')$ count requests (after local aggregation) with high probabiliy.

\noindent {\bf Balance}: Finally in the {\code Balance} stage each worker will send and receive at most $O(B')$ tuples even when they duplicate some tuples because we assume $B' \ge w$. 

In summary, each index lookup in each stage and the sending of a tuple in the {\code Balance} stage constitutes one unit of computation and communication. Therefore in each batch, with probability at least $1-O(\frac{1}{\IN \times MaxOut_Q})$, each worker  does $O(B')$ communication. Since the computation workers do is proportional to their communication, each worker, with the same probability do $O(B')$ computation. In the very beginning of the computation where the input is partitioned across workers, each worker has a load of $O(\frac{\mathrm{IN}}{w})$. Therefore the load of each worker is $O(\frac{\mathrm{IN}}{w} + B')$, completing the proof.
\end{proof}

\section{Comparison With Shares} 
\label{app:wco-load}
We compare the costs of BiGJoin-S with the variant of the Shares algorithm in reference~\cite{koutris:worst-case-parallel}, called {\em the Hypercube algorithm}, which is the most efficient version of the Shares algorithms in literature for arbitrary inputs. We start by a comparison of the total communication costs of Hypercube and BiGJoin-S. 

The Hypercube algorithm from reference~\cite{koutris:worst-case-parallel} is worst-case communication optimal for a subclass of queries but not every query. Recall that in this paper, we say an algorithm's communication cost is worst-case optimal if under any parallelism level $w$, its communication is asymptotically $O(MaxOut_Q)$. We first explain the class of queries on which the Hypercube algorithm from the reference is worst-case optimal. Assume that the input relations of a query $Q$ are the same size. We define two query-specific parameters, $\rho^*$ and $\tau^*$, to explain these class of queries. First parameter is the {\em fractional edge cover number} $\rho^*$. A {\em fractional edge cover} gives a non-negative weight $e_i$ to each relation $R_i \in Q$ such that for every attribute $a_j$, the weights of the relations that contain $a_j$ is at least 1. The minimum of fractional edge covers for a query is $\rho^*$. When sizes of input relations are the same, the AGM bound can be shown to be equivalent to $\Omega(\mathrm{IN}^{\rho^*})$.  The second parameter is the {\em fractional edge packing number} $\tau^*$. A {\em fractional edge packing}, similar to fractional edge cover, associates a non-negative weight $w_i$ to each relation except this time the sum of the weights of relations containing every attribute $a_j$ should be at most $1$. The maximum of fractional edge packing number is $\tau^*$. These quantities are not duals of each other and are not necessarily related. That is for some queries $\tau^* < \rho^*$, for others $\rho^* < \tau^*$, and for others $\rho^* = \tau^*$. The Hypercube algorithm is worst-case optimal on queries when $\tau^* \le \rho^*$. Examples include the triangle query where $\tau^*=\rho^*=3/2$. However it is not worst-case optimal when $\rho^* < \tau^*$, except on queries in which a single relation $R_i$ contains all of the $m$ attributes in $Q$. We give an example to demonstrate this, which requires us to translate the load parameter of MPC to total communication cost.

Consider the query $Q=R(a_1, a_2, a_3), S_1(a_1, b_1), S_2(a_2, b_2),$ \linebreak $S_3(a_3, b_3), T(b_1, b_2, b3)$. This query has $\rho^*=2$ and $\tau^*=3$. The load of the Hypercube algorithm on $Q$ is $\Omega(\mathrm{IN}/w^{1/3})$. The algorithm runs a constant number of rounds so the total communication cost of the algorithm is $\Omega(w^{2/3}\mathrm{IN})$. This cost for small $w$ is of course better than worst-case optimal. 
\vspace{3pt} 
For example when $w=O(1)$, it's linear in the size of the input, which is the least communication any algorithm has to incur. However as $w$ increases the communication increases beyond the AGM bound of $IN^2$. As a simple example, consider running at a parallelism level when the load of the Hypercube algorithm is $L=\mathrm{IN}/w^{1/3}=O(1)$, implying that $w=IN^3$. Therefore the total communication cost $\Omega(w^{2/3}\mathrm{IN} = IN^3)$, which is more than BiGJoin-S's communication cost of $O(mn\mathrm{IN}^2)$, and this cost is independent of the number of workers. Therefore the Hypercube algorithm incurs more communication than BiGJoin-S at high levels of parallelism.

For total memory cost, note that when executing on any parallelism level $w$, the Hypercube algorithm requires more cluster memory than BiGJoin-S. This is true even on queries where Hypercube is worst-case optimal, i.e., its load $L=\Omega(\mathrm{IN}/w^{1/\rho^*})$. Notice that this means in a single round, the $w$ workers get a total of $\Omega(w^{1-1/\rho^*}\mathrm{IN})$ data. Therefore the algorithm uses $L=\Omega(\mathrm{IN}/w^{1/\rho^*})$ memory per-worker and $\Omega(w^{1-1/\rho^*}\mathrm{IN})$ cluster memory, which is more than $O(mn\mathrm{IN}/w + mB')$ per-worker and $O(mn\mathrm{IN} +mB)$ cluster memory consumption of BiGJoin-S, when the batch size $B=wB'$ is picked small enough. In practice, we expect the batch sizes to be negligible compared to the $\Omega(\mathrm{IN})$ memory cost of indexing the relations and we need them to be logarithmic in the size of the input and the AGM bound to get our theoretical results with high probability. 

For load, or communication and memory per machine we make the following observation. Notice that by picking a small enough batch size, BiGJoin-S's load can always be made optimal and equal to $O(\mathrm{mnIN}/w)$ (up to the parameters $m$ and $n$, which only depend on $Q$ but not $\IN$). This shows that BiGJoin-S will scale linearly as we add more workers to the cluster. In contrast, the Hypercube algorithm requires $O(\mathrm{IN}/w^{1/\tau^*})$ or $O(\mathrm{IN}/w^{1/\rho^*})$ depending on the query, both of which are worse that BiGJoin-S's load, and does not scale linearly with increasing number of workers when $\tau^*, \rho^* > 1$.

Finally, for the number of rounds of computation the Hypercube algorithm performs better and runs a constant number of rounds while BiGJoin-S runs $O(\frac{mnMaxOut_Q}{B})$ rounds. 
\fi
\end{document}